\documentclass[aps,preprint,showpacs,superscriptaddress,groupedaddress]{revtex4}  
\usepackage{graphicx}  
\usepackage{dcolumn}   
\usepackage{bm}        
\usepackage{amssymb}   
\usepackage{amsmath}
\usepackage[colorlinks=true, citecolor=blue, urlcolor=blue, linkcolor = blue]{hyperref}
\usepackage{bm}
\usepackage{color}
\usepackage{bookmark}
\usepackage{tabularx}
\usepackage{mathtools}
\usepackage{microtype}
\usepackage{cancel}
\usepackage{relsize}
\usepackage{float}
\usepackage[caption = false]{subfig}
\usepackage{physics}

\begin{document}

\title{Q-ball mechanism of  electron transport and spin/lattice excitations properties of high-T$_c$ superconductors}
\author{S. I. Mukhin}
\affiliation{Theoretical Physics and Quantum Technologies Department, NUST MISIS, Moscow, Russia}
\date{\today}

\begin{abstract}
In this work the Q-ball mechanism of high-Tc superconductivity in cuprates, recently proposed by the author, is applied to explanation of the electron and spin/lattice properties found experimentally. Euclidean Q-balls (nontopological solitons) are Bose condensed collective spin/charge density wave fluctuations (SDW/CDW) with zero static mean that arise inside a Q-ball volume simultaneously with Cooper/local pairs superconducting condensate for which they serve as the 'pairing glue'. The Bose condensates of  SDW/CDW fluctuations possess wave-vectors Q$_{DW}$ that connect corresponding 'nested' regions of the Fermi surface near the van Hove singularities causing self-consistently pairing of the fermionic states that form superconducting  Bose condensate. The self-consistency equations for the quantum orders in the form of periodic in the imaginary Matsubara time SDW/CDW fields signify a new development beyond the Landau static mean-field order parameter approach. Due to Cooper/local pairs condensation the Q-balls possess lower total energy in comparison with not condensed thermal SDW/CDW fluctuations in the same volume. The Q-balls of finite volume exist due to conservation of the Noether "charge" Q caused by the breaking of chiral symmetry along the Matsubara time axis in the Euclidean space-time by the condensed SDW/CDW fields.  It is shown that Q-balls 'gas' is formed at temperature T$^*$ greater than bulk superconducting transition temperature T$_c$, both defined by inverse correlation length of the SDW/CDW field. The superconducting condensates inside the Q-balls induce a spectral gap on the nested parts of the Fermi surface in the 'pseudogap' phase of cuprates, where the Q-ball scenario was supported recently by micro X-ray diffraction data in HgBa$_2$CuO$_{4+y}$. It is found analytically that scattering of itinerant fermions on the Q-balls fluctuating fields causes linear temperature dependence of electrical resistivity in the interval of temperatures below T$^*$ and above the superconducting dome maximum T$_c$ , reminiscent of the famous 'Plankian' behavior in the 'strange metal' phase. Besides, scattering of phonons on the superconducting condensates inside the CDW Q-balls induces splitting of the in-plane phonon brunches into the softened and hardened ones close to Q$_{CDW}$ wave vectors in the Brillouin zone, while scattering of the spin wave excitations on the superconducting condensates inside the SDW Q-balls leads to the "hourglass" dispersion spectra close to the antiferromagnetic Q$_{SDW}$  wave vectors, both in correspondence with the neutrons scattering experiments. Analytically calculated contour plot of the Q-balls phase diagram, with superconducting dome boundary touching the upper 'strange metal' one, as well as the diamagnetic response of Q-balls gas above the T$_c$ dome, are in qualitative accord with experimental data in high-T$_c$ cuprates. 
\end{abstract}

\maketitle

\section{Introduction}
It was demonstrated previously \cite{Mukhin(2022), Mukhin(2022_1)} that electron/hole Fermi surface possessing 'nested' parts with high enough density of states , e.g. extended Van Hove singularity \cite{campuzano}, makes the system unstable to formation of the Euclidean Q-balls (nontopological solitons) that constitute Bose condensed collective spin/charge density wave fluctuations (SDW/CDW) with zero static mean that arise inside a Q-ball simultaneously with Cooper/local pairs superconducting condensate for which they serve as the 'pairing glue'. Bose condensed collective spin/charge density wave fluctuations (SDW/CDW) possess the wave-vectors Q$_{DW}$ that connect corresponding 'nested' regions of the Fermi surface near the van Hove singularities, e.g. in the doped cuprates. Consideration of  the self-consistency equations for the quantum order parameters in the form of periodic in the imaginary Matsubara time SDW/CDW semiclassical fields with zero mean actually signifies a new development beyond the Landau static mean-field order parameter approach. Corresponding free energy of the system is introduced then using Euclidean action instead of Landau free energy functional. In a particular case considered below the Bose condensation of the Matsubara time dependent SDW/CDW collective fluctuations forming a Q-ball is accompanied self-consistently by formation of the Cooper/local pair superconducting condensate formed of the $Q_{DW}$ - 'nested' fermionic states, thus making Q-ball formation energetically favourable. The Q-ball SDW/CDW semiclassical field with $U(1)$ symmetric bare Hamiltonian  rotates with Matsubara bosonic frequency $\Omega=2\pi T$, thus, breaking chiral symmetry along the Matsubara time axis in Euclidean space-time and possessing zero mean. This rotation provides conservation of the Noether "charge" Q proportional to the number of elementary SDW/CDW excitations (spin-waves/phonons) condensed inside the Q-ball. The extra conserved quantity Q (besides the energy) makes the Q-ball volume finite. The temperature T$^{*}$ of  the Q-balls condensation is proportional to the inverse correlation length (that defines the fluctuation's 'mass') of the SDW/CDW field and is much higher than static bulk SDW/CDW transition temperature \cite{Mukhin(2022_1)}. The latter falls down below the dome of  superconducting phase transition temperatures T$_c$ defined by the Q-ball induced Cooper/local pairing mechanism of superconductivity \cite{Mukhin(2022_1)}. The quintessence of  why T$^{*}$ and T$_c$ are relatively high with respect to the Bardeen-Cooper-Schrieffer (BCS) predicted superconducting transition temperature $T^{BCS}_c$ lies in that  T$^{*}$ and T$_c$ characterise instability with respect to condensation of  the semiclassical Q-ball SDW/CDW field coherently mediating Cooper/local pairing of the 'nested' fermionic states inside the Q-balls. This picture is drastically different from the BCS scenario of  condensation of  the Cooper pairs due to coupling of fermions via incoherent gas of single spin waves/phonons. Opening of the pairing energy gap  in the fermionic spectrum of the Q-balls on the 'nested' parts of the Fermi surface simultaneously transforms the whole system into 'pseudogap' (PG) phase. The condition that the Q-balls field solutions minimise the free energy of the whole system provides the self-consistency equations solved in \cite{Mukhin(2022), Mukhin(2022_1)}, that extend the Ginzburg-Landau static order parameter approach. Previously real and imaginary parts of 'rotating' in real time order parameter field, that constituted two scalar fields with nonderivative interactions in Minkowski space-time, were considered by Sidney Coleman as a Q-ball ansatz for baryons \cite{Coleman (1985)} in the supersymmetric standard model, where conserved Noether charge Q counted the number of baryons associated with the ${U}(1)$ symmetry of the squarks field \cite{Rosen, Coleman (1985), Lee and Pang (1992)}. The present paper extends farther theoretical investigation of the Euclidean Q-ball model predictions to the electron transport, phonon/spin dispersions splitting and diamagnetic properties of high-T$_c$ superconductors above T$_c$. In particular, the T-linear temperature dependence of  electrical resistivity in the interval of temperatures $T_{c} < T < T^*$  above superconducting dome maximum T$_c$ is derived analytically. The diamagnetic behaviour observed experimentally in the "strange metal" phase \cite{Niven, Li} is derived analytically as well. 
Here we also investigate the changes in the spectra of the quantum spin and phonon excitations due to their scattering on the superconducting condensates inside the Q-balls. In particular, magnetic and phonon dispersions, analytically derived below, demonstrate direct correspondence with the experimentally found famous 'hourglass' spin-waves dispersion in the vicinities of the SDW antiferromagnetic wave-vectors \cite{Tranquada_2014} and splitting into hardened and softened modes of the optical phonons found in the inelastic neutron scattering studies \cite{Egami, Keimer} in the high-T$_c$ cuprates in the vicinities of the CDW  'nesting' wave vectors of the Fermi surface regions in the PG phase.
It is important to stress here, that unlike in the usual static 'nesting' scenario with resulting long-ranged SDW/CDW ordering competing with superconducting pairing \cite{Tranq1, MatMuk}, the Q-ball semiclassical fields possess zero static mean values and therefore may belong to the 'local hidden orders'. Hence, the Q-ball induced magnetic and phonon spectral anomalies differ from the conventional Peierls-like pictures of static SDW/CDW formation, where electron-phonon/magnon coupling drives to zero corresponding dispersion branches of the excitations at the ordering wave vector of a static (spin)lattice distortion. Inelastic x-ray scattering studies of the phonon softening in high-T$_c$ cuprates were also considered in relation with phonon coupling to CDW fluctuations, see \cite{Pepin, Seibold, Caprara} and references therein. Previously, the Q-balls theory predictions for X-ray scattering \cite{Mukhin(2023)} were found in favourable accord with experimental results of X-ray diffraction in high-Tc cuprates superconductors in the heterogenous/PG phase \cite{campi22, campi, Annette}. 

The plan of the paper is as follows. In the next Sections II  and III a quintessence of  the Euclidean Q-balls picture is presented. Analytical derivations of the major parameters of  the Q-balls, including temperature dependence of their energy, semiclassical Q-ball field amplitude  and related phase diagram of the system possessing Q-balls are presented. In particular, it is demonstrated that the first order transition temperature T$^*$ is proportional to the inverse correlation length of the short-range spin/charge density wave fluctuations.  An idea of a Matsubara time-dependent semiclassical `pairing glue' between fermions in cuprates, but for an itinerant case, was proposed earlier \cite{Mukhin (2018)}. Described here superconducting pairing mechanism mediated by semiclassical Q-ball field is distinct from the usual phonon- \cite{BCS, elis} or spin-fermion coupling models \cite{Chubukov (2003)} considered previously for high-T$_c$ cuprates, based upon exchange with infinitesimal  spin- and charge-density excitations \cite{Seibold} or polarons \cite{Bianconi (1994)} in the usual Fr\"{o}hlich picture. In section IV the self-consistently determined temperature dependence of the Q-ball field amplitude is used for an analytical derivation of the $T$-linear temperature dependence of  the inverse life-time of fermionic excitations due to scattering on Q-balls in the normal phase. This inverse life-time $T$-linear temperature dependence is in accord with experimentally measured linear temperature dependence of electrical resistivity in the 'strange metal' phase \cite{Niven}. In Section V the paraconductivity calculation method by Alex Abrikosov \cite{Alex} is used  for derivation of electrical resistivity dominated by Q-balls slide. In Section VI diamagnetic response of Q-balls 'gas' is calculated and qualitative accord with experimental data of  L. Li et al. \cite{Li} is found. In Section VII the 'hourglass' like dispersion in the vicinity of antiferromagnetic SDW wave vectors of  the magnetic excitations, and splitting of the Cu-O in-plane phonon brunches  into the softened and hardened ones close to CDW fluctuations wave vectors in the Brillouin zone are derived analytically and plots generated by Wolfram Mathematica program are presented. Conclusions follow in Section VIII. 

\section{Euclidean extension of Landau's free energy functional approach}
This section contains an overview of the previous work on Q-balls in Euclidean space-time that helps a reader to navigate from the origins of the main expressions to the derivation of  the temperature dependences of the field amplitude used to achieve the main results of the present paper. More detailed information is incorporated in the works published previously \cite{Mukhin(2022), Mukhin(2022_1), Mukhin(2023), Annette}. 

The usual Hamiltonian describing interaction between electrons and phonons/spin waves is written in the form \cite{AGD}:
\begin{eqnarray}
\begin{split}
H =\sum_{\vec{p},\sigma}\varepsilon_{\vec{p}}c^{\dagger}_{\vec{p},\sigma}c_{\vec{p},\sigma} + \sum_{\vec{q}}\omega_{\vec{q}}b^{\dagger}_{\vec{q}}b_{\vec{q}}+g\sum_{\vec{p},\vec{q},\sigma}\left(\dfrac{\omega_{\vec{q}}}{2V}\right)^{1/2}\left(c^{\dagger}_{\vec{p},\sigma}c_{\vec{p}+\vec{q},\sigma}b^{\dagger}_{\vec{q}}+h.c.\right)
 \end{split}
  \label{Hepf}
  \end{eqnarray}
\noindent leading to the effective interaction Hamiltonian used by Eliashberg \cite{elis} in the Fr\"{o}hlich picture \cite{AGD} of superconductivity mechanism:
\begin{eqnarray}
H_{int}(\tau)=\dfrac{g^2}{V}\sum_{{\bf{p,p',q}},\sigma,\sigma'}\int_0^{\beta}d\tau' D_{{\bf{q}}}(\tau-\tau')\left(c^{+}_{{\bf{p}},\sigma}(\tau)c_{{\bf{p+q}},\sigma}(\tau)c^{+}_{{\bf{p'+q}},\sigma'}(\tau')c_{{\bf{p'}},\sigma'}(\tau')\right)
\label{intEl}
\end{eqnarray}
\noindent where phonon/spin-wave propagator $D_{{\bf{q}}}(\tau-\tau')$ equals \cite{AGD}:
 
 \begin{eqnarray}
&D_{{\bf{q}}}(\tau)=-\dfrac{\omega_{{\bf{q}}}}{2}\left[\left(N({\bf{q}})+1\right)\exp{\left\{\omega_{{\bf{q}}}|\tau|\right\}} +
N({\bf{q}})\exp{\left\{-\omega_{{\bf{q}}}|\tau|\right\}}\right];\nonumber \\
&N({\bf{q}})=\left[\exp{\left\{\omega_{{\bf{q}}}/T\right\}}-1\right]^{-1}
\label{Dph}
\end{eqnarray}
\noindent where the Boltzmann constant is taken for unity $k_B=1$. The Q-ball picture of superconductivity proposed recently \cite{Mukhin(2022), Mukhin(2022_1), Mukhin(2023)} considers possibility of Bose-condensation of phonons/spin-waves in some particular points $\left\{\Omega,{\bf{q}}\right\}$ of their $D+1$ Euclidean frequency-momentum phase-space thus leading to a substitution of the $\left(\dfrac{\omega_{\vec{q}}}{2V}\right)^{1/2}b^{\dagger}_{\vec{q},\Omega}\;,\;\; \left(\dfrac{\omega_{\vec{q}}}{2V}\right)^{1/2}b_{\vec{q},\Omega}$ creation/annihilation phonon/spin-wave Bose-operators with the Matsubara time - dependent order parameter represented by the Bose-condensate c-number operators $\zeta^{+}_{0},\zeta_{0}$:
\begin{eqnarray}
\zeta^{+}_{0}={M^{*}}_{\vec{q},\Omega}\exp{\left\{-i{\bf{q}}\cdot{\bf{r}}+i\Omega\tau\right\}}\,;\; \zeta_{0}=M_{\vec{q},\Omega}\exp{\left\{i{\bf{q}}\cdot{\bf{r}}-i\Omega\tau\right\}}
 \label{BoM}
  \end{eqnarray}
\noindent
 describing a finite density of the corresponding lattice charge/spin density distortion:  $\zeta^{+}_{0}\zeta_{0}=|M_{\vec{q},\Omega}|^2$. Importantly, the Bose c-number operators in Eq. (\ref{BoM}) possess chirality along the Matsubara time axis, i.e. the symmetry $\Omega\rightarrow -\Omega$ is broken. This is necessary condition for nonzero conserved Noether "charge" Q to exist along the winding trajectory of the Bose-condensate in Euclidean space leading to a finite Q-ball volume. It is shown in \cite{Mukhin(2022), Mukhin(2022_1), Mukhin(2023)} that the free energy of the system may possess a minimum with respect to $|M_{\vec{q},\Omega}|$ amplitude and Q-ball volume $V_Q$. Such a nontrivial minimum arises for a finite volume Q-balls due to condensation of superconducting Cooper/local fermionic pairs, for which the operators $\zeta^{+}_{0},\zeta_{0}$ play the role of the 'pairing glue' bosons. Thus, this new mechanism of superconductivity describes the picture of fluctuating semiclassical  charge/spin-density waves with a particular Bose frequency $\Omega=2\pi nT$ inside the Q-balls, represented by operators $\zeta^{+}_{0},\zeta_{0}$, that are not at all competing with superconductivity, but on the contrary, are creating it in a self-consistent manner.   To see how this happens, one may start from a simple t-U Hubbard Hamiltonian: 

\begin{eqnarray}
\begin{split}
H =-t\sum_{\langle i,j\rangle,\sigma}c^{\dagger}_{i,\sigma}c_{j,\sigma} + U\sum_i  \hat{n}_{i, \uparrow}\hat{n}_{i, \downarrow} - \mu \sum_{i, \sigma} \hat{n}_{i, \sigma} ,
 \end{split}
  \label{Hub}
  \end{eqnarray}
\noindent and use formally the Hubbard-Stratonovich decoupling procedure with a scalar complex field $M_{{\bf{q}}}(\tau,{\bf{r}})=\zeta_{0}$ defined in Eq. (\ref{BoM}): $M_{{\bf{q}}}(\tau,{\bf{r}})=M_{{\bf{q}},\Omega}\exp{\left\{i{\bf{q}}\cdot{\bf{r}}-i\Omega\tau\right\}}$, that leads either to 
electrons/holes scattering on SDW field $\sigma M_{{\bf{q}}}(\tau,{\bf{r}})$:  
\begin{align}
H^{SDW}_{int}(\tau)=\sum_{{\bf{q,Q}},\sigma}\left(c^{+}_{{\bf{q+Q}},\sigma}(\tau)M_{\bf{Q}}(\tau)\sigma c_{{\bf{q}},\sigma}(\tau)+H.c.\right)
\label{intinstS}
\end{align}
\noindent or to electrons/holes scattering on CDW field $M(\tau,{\bf{r}})$ in the crystal lattice: 
\begin{align}
H^{CDW}_{int}(\tau)=\sum_{{\bf{q,Q}},\sigma}\left(c^{+}_{{\bf{q+Q}},\sigma}(\tau)M_{\bf{Q}}(\tau)c_{{\bf{q}},\sigma}(\tau)+H.c.\right)
\label{intinstC}
\end{align}
\noindent where the shorthand $M_{\bf{Q}}(\tau)\equiv M_{{\bf{Q}},\Omega}\exp{\left\{-i\Omega\tau\right\}}$, and $\sigma$ spin factor is missing in the charge - fermion coupling vertex $c^{+}Mc$ and interaction representation is  implied for the fermionic creation/annihilation operators with the hamiltonian (\ref{Hub}), but without $U$-term. The Hubbard-Stratonovich field of a Q-ball nontopological soliton $M(\tau,{\bf{r}})$ is sought for in the form:

\begin{align}
M(\tau,{\bf{r}})=e^{i{\bf{Q\cdot r}}-i\Omega\tau}M\Theta\left\{{\bf{r}}\right\}\;;\quad \Theta({\bf{r}})\equiv\begin{cases}1 ;\;{\bf{r}}\in V ;\\
0;\;{\bf{r}}\notin V.\end{cases}\;, \Omega=2\pi nT,\; n=1,2,...
\label{step}
\end{align}
\noindent where $V$ is the Q-ball volume that minimises the Euclidean action found below. Thus, besides Matsubara time periodicity $M(\tau+1/T,{\bf{r}})=M(\tau,{\bf{r}})$, the field is assumed to break chirality along the Matsubara time axis.
A simple model Euclidean action ${S}_{M}$ for the scalar field $M(\tau,{\bf{r}})$, could be written as:
\begin{eqnarray}
&S_{M}^0&=\int_0^{\beta}\int_Vd\tau d^D{\bf{r}}\dfrac{1}{g}\left\{ |\partial_{\tau}M|^2 +s^2 |\partial_{{\bf{r}}}M|^2 + {\mu _0 ^2 }{|M|^2}\right\},\; \nonumber \\
&&M\equiv M(\tau,{\bf{r}}) \label{Eu}
\end{eqnarray}
\noindent where $s$ is bare propagation velocity, and the `mass' term $\mu_0^2\sim 1/\xi^2$ imposes finite correlation length $\xi$ of the fluctuations. Considering now  the Fermi surface with the 'nested' parts with high enough density of states, e.g. Van Hove singularity \cite{campuzano}, being connected by a wave vector ${\bf{Q}}_{DW}$, one would customary declare as the next step that field $M(\tau,{\bf{r}})$ is Matsubara time $\tau$ independent, with a 'nesting' wave-vector ${\bf{Q}}={\bf{Q}}_{DW}$ becoming a SDW or CDW wave vector formed under (Peierls-like) phase transition, that forms corresponding energy gap in the fermionic spectrum. Hence, superconductivity would be then considered as a 'competing order', see e.g. \cite{Tranq1}, \cite{MatMuk} for a recent overview. The Q-ball picture approach is drastically different. 

\subsection{Hubbard-Stratonovich Q-ball field as the 'pairing glue'} 
It was shown \cite{Mukhin(2022), Mukhin(2022_1)} that Euclidean action of the Hubbard-Stratonovich field may develop a semiclassical minimum of the Q-ball universality class (nontopological Euclidean soliton) (\ref{step}) when simultaneously the fermions that are 'decoupled' by this field in  Eqs. (\ref{intinstS}), (\ref{intinstC}) start locally condense into a Cooper/local pair superconducting condensate.
The soliton field  $M(\tau,{\bf{r}})$ is periodic in Matsubara time with  zero mean value, and, therefore, is called 'thermodynamic quantum time crystal' \cite{Efetov, Timur}. But, most important, besides Matsubara time periodicity $M(\tau+1/T,{\bf{r}})=M(\tau,{\bf{r}})$, the field is assumed to break chirality along the Matsubara time axis, thus taking the form in Eq. (\ref{step}). The Q-balls with 'counterclockwise' chiral combination $e^{i{\bf{Q_{DW}\cdot r}+\Omega\tau}}$ are also allowed as separate objects. After the Hubbard-Stratonovich SDW/CDW field $M(\tau,{\bf{r}})$ in the above form is inserted into decoupled parts of the t-U Hubbard Hamiltonian Eqs. (\ref{intinstS}) and (\ref{intinstC}) the Cooper paired fermions are integrated out creating an additional energy term $U_f$ in the effective action of the $M(\tau,{\bf{r}})$ field:
\begin{align}
&VU_{f}(|M(\tau,{\bf{r}})|)=\Delta\Omega_s=-T\ln\dfrac{Tr\left\{e^{-\int_0^{\beta}H_{int}(\tau)d\tau}{\cal{G}}(0)\right\}}{Tr\left\{{\cal{G}}(0)\right\}}\equiv \Omega_s-\Omega_0;\,{\cal{G}}(0)\equiv e^{-\beta H_0};\\
&H_0=\sum_{{\bf{q}},\sigma}\varepsilon_q c^{+}_{{\bf{q}},\sigma}c_{{\bf{q}},\sigma}\label{DOI} 
\end{align}
\noindent Here $H_0$ is Hamiltonian of the noninteracting fermions on a lattice with the bare dispersion $\varepsilon_q$ and $\Delta\Omega_s$ is the electron pairs contribution to the free energy. The latter is calculated via standard procedure \cite{AGD} and its result is presented in detail in \cite{Mukhin(2022), Mukhin(2022_1)}:
\begin{align}
&U_{f}=-\dfrac{4\nu \varepsilon_0\Omega}{3}I\left(\dfrac{M}{\Omega}\right)\;,\;M\equiv |M(\tau)|\label{UFO1}\\
&I\left(\dfrac{M}{\Omega}\right)= \int_\gamma^{M/\Omega}d\alpha\dfrac{\alpha \sqrt{2\alpha \left(\alpha -\gamma\right)}}{(1+8\alpha \left(\alpha -\gamma\right))}\tanh{\dfrac{\sqrt{2\alpha\left(\alpha-\gamma\right)}\Omega}{\varepsilon_0}}\tanh{\dfrac{\sqrt{2\alpha\left(\alpha-\gamma\right)}\Omega}{2T}}\,,\gamma\approx 1/2
\label{UFOI1}
\end{align} 
\noindent where $2\varepsilon_0$ signifies the width of the interval along the energy axis, $|\varepsilon_p-\mu|\leq \varepsilon_0$, with nonzero density of fermionic states $\nu(\varepsilon_p)\approx\nu$ obeying the 'nesting' relation $\varepsilon _{p - Q_{DW}}  =  - \varepsilon _p$ around the chemical potential $\mu$. Expression (\ref{UFOI1}) is remarkable, since it bears result of the important two step self-consistent 'bootstrap procedure' \cite{Mukhin(2022), Mukhin(2022_1)}. Namely, at the first step the following relation is derived using mentioned above procedure for the free energy calculation via integration over a variable coupling strength $\alpha$ \cite{AGD}: 
\begin{align}
&\dfrac{\partial\Omega_s}{\partial \alpha}=T{\int_0^{\beta}{\left\langle\dfrac{\partial H_{int}(\tau)}{\partial \alpha}\right\rangle} d\tau}=
-\frac{T}{\alpha}\int_0^{\beta}{\int}_0^{\beta}d\tau d\tau_1 \left\langle H_{int}(\tau) H_{int}(\tau_1)\right\rangle =
\nonumber\\
&-\frac{TV}{\alpha}|M|^2T\sum_{\omega,{\bf{p}},\sigma}\sigma\bar{\sigma}\overline{F}_{\sigma,\bar{\sigma}}(\omega,{\bf{p}}){F}_{\bar{\sigma},\sigma}(\omega-\Omega,{\bf{p}}-{\bf{Q_{DW}}})\alpha^2 \;,
\label{DOs}
\end{align}

\noindent where the  loop of Gor'kov anomalous functions $F^\dagger, F$ describing the condensed paired fermions contains the Q-ball semi-classical field propagator $D(\tau-\tau', {\bf{r}}-{\bf{r}}')\sim {M(\tau',{\bf{r}}')}^*\cdot M(\tau,{\bf{r}})$ instead of the usual phonon-/spin-wave propagator in the Fr\"{o}hlich picture \cite{AGD}. 

\subsection{Eliashberg-like equation with Q-ball field as the 'pairing  glue'}

The second step of the 'bootstrap' self-consistent procedure would be then to solve the Eliashberg-like equation for the Gor'kov anomalous fermionic Green's functions $F^\dagger, F$  with the Q-ball field (\ref{step}) playing role of the 'pairing  glue'  that couples 'nested' fermionic states on the bare Fermi surface \cite{Mukhin(2022), Mukhin(2022_1)}:
\begin{eqnarray}
F_{p,\sigma}(\omega)=-\Sigma _{2p,\sigma } (\omega )K_{p}(\omega)=-K_{p}(\omega)\left[T\sum_{\pm \Omega}{{\cal{D}}_{Q_{DW}} (\Omega )F_{p-Q_{DW},\sigma}(\omega -\Omega )}\right]\;,\label{Feq} 
\end{eqnarray}
where : 
\begin{eqnarray}
K_{p}(\omega)=\left\{|  i\omega-\varepsilon _{p } -\Sigma _{1p ,\sigma } (\omega)|^2+|\Sigma_{2p,\sigma}(\omega)|^2\right\}^{-1}\approx \left\{\omega^2 + \varepsilon _{p }^2 +|\Sigma_{2p,\sigma}(\omega)|^2\right\}^{-1}. 
\label{K0}
\end{eqnarray}
\noindent Here:
\begin{eqnarray}
{\cal{D}}_{Q_{DW}} (\Omega )\equiv \dfrac{M^2}{T}\,, \quad{\cal{D}}_{Q_{DW}} (\tau )\equiv 2M^2\cos{(\Omega\tau)}\,. \label{DM}
\end{eqnarray}
\noindent The self-energy function $\Sigma_{2p,\sigma}(\omega)$ is approximated with parabolic function of the bare fermionic dispersion $\varepsilon_p$ in the vicinity of the Fermi energy:
\begin{align}
|\Sigma_{2p,\sigma}(\omega)|^2=g_0^2-\varepsilon_p^2\,.
\label{sig2}
\end{align}
\noindent As it was found previously \cite{Mukhin (2018)}, compare \cite{elis}, the normal self-energy function $\Sigma _{1p ,\sigma } (\omega)$ leads to renormalisation of the Fermi velocity and the chemical potential shift and could be neglected. Then, using approximation (\ref{sig2}) for the self-energy function $\Sigma_{2p,\sigma}(\omega)$ it is straightforward to convert Eq. (\ref{Feq}) directly into differential equation of the Mathieu type \cite{Mukhin(2022), Mukhin(2022_1)}:

\begin{equation}
\partial^2_{\tau}F_{p,\sigma}\left( \tau  \right) +\left(2M^2\cos{\left( \Omega\tau  \right)} - g_0^2\right)F_{p,\sigma} \left( \tau  \right)=0,\;\quad F_{p,\sigma} \left( {\tau}+\dfrac{1}{T} \right)=-F_{p,\sigma} \left( {\tau} \right)\;, \Omega = 2\pi nT\,,
\label{Mathieu}
\end{equation} 
\noindent exact solutions of which are well known, and the anti-periodicity condition \cite{AGD} for the sought for solution for the fermionic Green's function $F_{p,\sigma} \left( {\tau} \right)$  is explicitly indicated.  The lowest possible eigenvalue $-g_0^2$ is then \cite{Witteker}:
\begin{equation}
g_0^2 = 2M\left(M-\gamma\Omega\right); \quad \gamma\approx 1/2\,,
\label{Mg00}
\end{equation}
\noindent Hence, a non-zero anti-symmetric solution $F(\tau)$ of Eq. (\ref{Mathieu}) with real eigenvalue $g_0^2$ exists when $M>\gamma \Omega$. This result is remarkable: it follows from relations (\ref{Feq}), (\ref{sig2}) and (\ref{Mg00}),  that Q-ball of 'glue boson' condensate possessing amplitude $M>\gamma\Omega$ also possesses a Cooper pair condensate characterised with non-zero anomalous Green's function $F_{p,\sigma}$: 

\begin{eqnarray}
F_{p,\sigma}(\omega)=-\Sigma _{2p,\sigma } (\omega )K_{p}(\omega)\approx-\dfrac{\sqrt{g_0^2-\varepsilon_p^2}}{\omega^2 + g_0^2} ,
\label{Fam} 
\end{eqnarray}

\noindent Since non-zero solution for the superconducting gap arises starting from finite amplitude $M>\gamma \Omega$, the Q-ball phase transition is of the first order with respect to the Q-ball field amplitude $M$. On the other hand, at finite $M$, that obeys: $M=\gamma\Omega$ the density of superconducting condensate inside the Q-ball equals zero. Hence the Q-ball phase transition is of the second order with respect to the density of superconducting condensate inside the Q-ball. Next,  at the last step, an amplitude $M$ is substituted with $\alpha M$ in (\ref{Mg00}) according to the definition of the formal dimensionless coupling  parameter $\alpha$ in Eq. (\ref{DOs}), thus, leading to the final expression in Eqs. (\ref{UFO1}), (\ref{UFOI1}):

\begin{equation}
g_0^2(\alpha)= 2M\alpha\left(M\alpha-\gamma\Omega\right)\,,
\label{Mg0}
\end{equation}
\noindent which is used in the final 'bootstrap' expression for the effective Q-ball field energy $U_f$ in Eq. (\ref{UFOI1}). Hence, now one obtains the effective Euclidean action of the field  $M(\tau,{\bf{r}})$, that includes contribution from condensed superconducting fermionic pairs, $U_{f}(|M(\tau,{\bf{r}})|)$. The action has to be  searched for the assumed Q-balls quasi-classical minima:
 \begin{eqnarray}
 S_{M}=\int_0^{\beta}\int_Vd\tau d^D{\bf{r}}\dfrac{1}{g}\left\{ |\partial_{\tau}M|^2 +s^2 |\partial_{{\bf{r}}}M|^2 + {\mu _0 ^2 }{|M|}^2+ gU_{f}(|M|^2)\right\},\; M\equiv M(\tau,{\bf{r}})\,, \label{Eu1}
\end{eqnarray}
\noindent where $g=U^3V$, $V$ - is the system volume, $\mu_0^2\propto U^2$. Under these values choice the action (\ref{Eu1}) would indeed serve to decouple the interaction term $U\sum_i  \hat{n}_{i, \uparrow}\hat{n}_{i, \downarrow}$ in the t-U Hubbard Hamiltonian in Eq. (\ref{Hub}), that results then in appearance of the term $c^{+}_{{\bf{q+Q}},\sigma}(\tau)M_{\bf{Q}}(\tau)\sigma c_{{\bf{q}},\sigma}(\tau)$ or similar in the $H_{int}$ Hamiltonians in  Eqs. (\ref{intinstS}),(\ref{intinstC}). Hence, the amplitude of the field  $M(\tau,{\bf{r}})$ relates with the local condensed spin/charge densities as: $|\hat{n}_{i, \uparrow}\pm\hat{n}_{i, \downarrow}|\sim M(\tau,{\bf{r}})/U$. This in turn imposes upper limit on the amplitude of the field: $|M(\tau,{\bf{r}})|< U$ allowing for the  value of a local electron spin/charge that might condense in the SDW/CDW Q-ball fluctuation considered here \cite{Louk}. 
The model (\ref{Eu1}) is $U(1)$ invariant under the global phase rotation $\phi$: $M\rightarrow Me^{i\phi}$. Hence, corresponding  `Noether charge' is conserved along the Matsubara time axis \cite{Mukhin(2022), Mukhin(2022_1)}:
\begin{eqnarray}
Q=\int_V j_{\tau}d^D{\bf{r}}=\int_V\frac{i}{2}\left\{M^{*}(\tau,{\bf{r}})\partial_\tau M(\tau,{\bf{r}})-M(\tau,{\bf{r}})\partial_\tau M^{*}(\tau,{\bf{r}})\right\}d^D{\bf{r}}=\Omega M^2V\; , \label{Q}\; 
\end{eqnarray}
\noindent The `Noether charge' conservation causes occurrence of Matsubara time periodic, finite volume Q-ball semiclassical solutions, that otherwise would be banned in D$>2$ by Derrick theorem~\cite{Derrick} in the $\tau$-independent static case. Indeed, substitution of relation for the charge $Q$ (\ref{Q}) into Euclidean action (\ref{Eu1}) leads to the expression that possesses minimum at finite Q-ball volume $V_Q$:

\begin{eqnarray}
 S_{M}=\frac{1}{gT }\left\{\dfrac{Q^2}{VM^2}+ V[\mu _0 ^2M^2 + gU_{f}]\right\}, \label{SMQ}
\end{eqnarray}
\noindent  Provided the $\propto V$ term above is positive, the action $S_M$ is minimised by Q-ball volume $V_Q$:  
\begin{eqnarray}
 V_Q= \dfrac{Q}{M\sqrt{\mu _0 ^2 M^2+ gU_{f}(M)}}\;; \label{VQ}
 \end{eqnarray}
 \noindent for which the free energy of the Q-ball field equals:
 \begin{eqnarray}
 E_Q=TS^{min}_M=\dfrac{2Q\sqrt{\mu _0 ^2 M^2+ gU_{f}(M)}}{gM}=\dfrac{2Q\Omega}{g}, \label{aQ}
\end{eqnarray}
\noindent After cancellation of Q-factor in the last relation in Eq. (\ref{aQ}) above the following self-consistency equation emerges  \cite{Mukhin(2022_1)}:
\begin{eqnarray}
 0=(\mu _0 ^2 -\Omega^2)M^2+ gU_{f}(M). \label{self}
\end{eqnarray}
\noindent One then substitutes $U_{f}(M)$ from Eqs. (\ref{UFO1}), (\ref{UFOI1}) into self-consistency equation  Eq. (\ref{self}) and finds:
\begin{eqnarray}
(\mu _0 ^2-\Omega^2){M}^2 - \dfrac{4\Omega g \nu \varepsilon_0}{3}I\left(\dfrac{M}{\Omega}\right)  =0. \label{self1}
\end{eqnarray}
\noindent The above self-consistency equation was investigated in \cite{Mukhin(2022), Mukhin(2022_1)} numerically. Below, using approximate expression for the potential energy ${U}_{f}$ the $M(T)$ dependences will be found analytically and used for calculation of the temperature dependences of  electrical resistivity and diamagnetic response of the Q-ball gas above superconducting transition temperture T$_c$. Simultaneously, condition $E_Q=0$ at $Q\neq 0$ marks transition to the bulk superconducting state since the Q-ball volume $V_Q$ then diverges according to Eq. (\ref{VQ}). One important observation is in order. Expression (\ref{self1}) could be obtained in two different cases. Namely, there are two possibilities for the self-energy function $\Sigma _{2p,\sigma }$: the $d$-wave symmetric behaviour of superconducting order parameter $\Sigma _{2p - Q_{DW},\sigma }  =  - \Sigma _{2p,\sigma }$ and the s-wave behaviour of it $\Sigma _{2p - Q_{CDW},\sigma }  =  \Sigma _{2p,\sigma }$. The first case realises when  spin (SDW) fluctuations couple to the fermions via interaction Hamiltonian (\ref{intinstS}), while the second case realises when charge (CDW) fluctuations instead of spin fluctuations couple to the fermions via interaction Hamiltonian (\ref{intinstC}). The particular choice of the two symmetries:  $\Sigma _{2p - Q_{DW},\sigma }  = \pm \Sigma _{2p,\sigma }$, is governed by the demand that contribution to the free energy in (\ref{DOs}) due to pairing would be negative. The $\sigma$ spin factor is missing in the charge - fermion coupling vertex $c^{+}Mc$ in (\ref{intinstC}). This leads to the absence of the factor $\sigma\bar{\sigma}=-1$ in the Eq. (\ref{DOs}) in the CDW Q-ball field case. Hence, in order to keep $U_f <0$, as is necessary for the Q-ball formation, one has to compensate for this sign change in the CDW Q-ball field case by the change of the sign of the Green's functions product $\overline{F}_{\sigma,\bar{\sigma}}(\omega,{\bf{p}}){F}_{\bar{\sigma},\sigma}(\omega-\Omega,{\bf{p}}-{\bf{Q_{DW}}})$ in Eq. (\ref{DOs}). Then, allowing for the structure of the Gor'kov's anomalous Green's function in Eqs. (\ref{Feq}), (\ref{K0}) one concludes, that relation between the values of superconducting order parameter in the points connected by the 'nesting' wave vector $Q_{CDW}$  should be altered with respect to Q-ball of SDW fluctuation, i.e in case of CDW Q-ball mediated pairing the 'nesting' wave vector should couple points with the same sign of superconducting order parameter.

\section {The phase diagram of the Q-balls gas}

Summarising,   Eq. (\ref{Eu1}) was used to describe effective theory of  the Fourier components of  the leading Q-ball (i.e., short-range) SDW/CDW fluctuations. Explicit expression for $U_{f}(|M(\tau,{\bf{r}})|)$ was derived and investigated in detail previously~\cite{Mukhin(2022), Mukhin(2022_1)}  by integrating out Cooper/local-pairs fluctuations in the `nested' Hubbard model with charge-/spin-fermion interactions. As a result, Q-ball self-consistency Equation (\ref{self}) was solved and investigated, and it was established that  Euclidean Q-balls  describe stable semiclassical short-range charge/spin-ordering fluctuations of finite energy that appear at finite temperatures below some temperature T$^*$, found to be $T^*= \mu_0/2\pi$ \cite{Mukhin(2022), Mukhin(2022_1)}. Next, it was also found that transition into pseudogap phase at the temperature T$^*$ is of the 1st order with respect to the amplitude $M$ of the Q-ball  SDW/CDW fluctuation and of 2nd order with respect to the superconducting gap $g_0$.  In particular, the following temperature dependences of these characteristics of the Q-balls were derived from Equations (\ref{UFOI1}), (\ref{Mg00}) and (\ref{self})  in the vicinity of  the transition temperature T$^*$ into Q-ball phase \cite{Mukhin(2022_1)} for the CDW/SDW amplitude: 
\begin{eqnarray}
M=\dfrac{\Omega}{2}\left(1+ \left( \dfrac{\left(T^*-T\right)15\sqrt{2}\mu _0}{g\nu}\right)^{\frac{2}{5}}\right),\quad T^*=\dfrac{\mu_0}{2\pi}\,,
 \label{Mstar}
\end{eqnarray}
 \noindent  and for the pseudogap $g_0$:
\begin{equation}
g_0^2= \left( {T}^*-T\right)^{\frac{2}{5}}{\Omega}^2\left(\dfrac{15\mu _0}{4g\nu}\right)^{\frac{2}{5}},
\label{g0star}
\end{equation}
\noindent which follows after substitution of Equation~(\ref{Mstar}) into Equation~(\ref{Mg00}). The above results differ from those presented in \cite{Mukhin(2022_1)} by the change of  the value of  $\gamma =1$, used previously, for the more precise value $\gamma =1/2$ in (\ref{Mg00}).

An expression for the fermionic pairing contribution to the potential energy, $U_{f}(|M|^2)$, found in Eqs. (\ref{UFO1}) and (\ref{UFOI1}) could be approximated away from the $T^*$ temperature in the form:
\begin{equation}
gU_{f}(M)=-\dfrac{4g\nu \varepsilon_0\Omega}{3}I\left(\dfrac{M}{\Omega}\right)\approx -\dfrac{g\nu \varepsilon_0}{3\sqrt{2}}\left(M- \dfrac{\Omega}{2}\right)\,,
\label{UFOA}
\end{equation} 
\noindent where now $\gamma\approx 1/2$ is used for the superconducting gap function $g_0$ :
\begin{equation}
g_0^2 \approx 2M\left(M-\dfrac{\Omega}{2}\right).
\label{Mg001}
\end{equation}
\noindent Hence, substituting the above expression into the pairing-induced effective potential energy of SDW/CDW field in Eq. (\ref{UFO1}), one now finds:

\begin{align}
&U_{eff}(M)=\mu _0 ^2 M^2+gU_f\approx \mu _0 ^2 M^2 -\dfrac{g\nu \varepsilon_0}{3\sqrt{2}}\left(M- \dfrac{\Omega}{2}\right)\equiv\nonumber\\
&\equiv\mu _0 ^2\left\{\left(M-\dfrac{g\nu \varepsilon_0}{6\sqrt{2}\mu _0 ^2}\right)^2+\dfrac{g\nu \varepsilon_0}{6\sqrt{2}\mu _0 ^2}\left(\Omega-\dfrac{g\nu \varepsilon_0}{6\sqrt{2}\mu _0 ^2}\right)\right\}.
\label{UFOI2}
\end{align} 
\noindent Then, one finds the value of the highest superconducting transition temperature $T_c=\Omega_c/2\pi$  from the Q-ball volume divergence condition in Eq. (\ref{VQ}):
\begin{align}
&U_{eff}(M)=\mu _0 ^2 M^2+gU_f=0=\mu _0 ^2\left\{\left(M-\dfrac{g\nu \varepsilon_0}{6\sqrt{2}\mu _0 ^2}\right)^2+\dfrac{g\nu \varepsilon_0}{6\sqrt{2}\mu _0 ^2}\left(\Omega_c-\dfrac{g\nu \varepsilon_0}{6\sqrt{2}\mu _0 ^2}\right)\right\}.
\label{UFOIC}
\end{align} 
\noindent From where it is straightforward to find:
\begin{align}
T_{c} =\dfrac{\Omega_c}{2\pi}= \dfrac{g\nu \varepsilon_0}{12\pi\sqrt{2}\mu _0 ^2}\,;\;\; M_c\equiv M(\Omega_{c})=\dfrac{g\nu \varepsilon_0}{6\sqrt{2}\mu _0 ^2}\equiv \Omega_{c}
\label{UFOIS}
\end{align} 
Now, substituting results from Eq. (\ref{UFOIS}) into Eq. (\ref{Mg001}) one finds directly the following relation between the superconducting gap $g_c$ and the temperature $T_c$:
\begin{equation}
g_c= \sqrt{ 2M_c\left(M_c-\dfrac{\Omega_c}{2}\right)}=\Omega_c\,;\;\;\dfrac{2g_c}{T_c}\equiv 2\pi\dfrac{2g_c}{\Omega_c}=4\pi\approx 12,57\,.
\label{Mg00S}
\end{equation}
\noindent The above results in Eqs. (\ref{UFOIS}), (\ref{Mg00S}) are remarkable from two points of view. First, the ratio $\approx 12,57$ of reduced superconducting gap in Eq. (\ref{Mg00S}) drastically differs from the BCS ratio $=3.5$ \cite{BCS} and compares batter with $\approx 7.4$ found for BSCCO high-$T_c$ compounds \cite{Fischer}. Second, expression for $T_c$ in Eq. (\ref{UFOIS}) permits, in principle, to infer relation with the isotope effect, since besides the product $g\nu \varepsilon_0$  the expression contains also the SDW/CDW characteristic parameter $\sim\mu _0^{-2}$. Simultaneously, the pseudogap/strange metal transition temperature $T^*= \mu_0/2\pi$ possesses the same parameter $\mu _0$ in the numerator. Hence, the isotope effects for $T^*$ and $T_c$ would contain exponents of the opposite signs related with Q-ball parameter $\mu _0$. 
The next application of the approximate expression for the fermionic pairing contribution to the potential energy, $U_{f}(|M|^2)$, found in Eq. (\ref{UFOA}) is even more impressive. 
\begin{figure}[H]
    \includegraphics[width=0.45\linewidth]{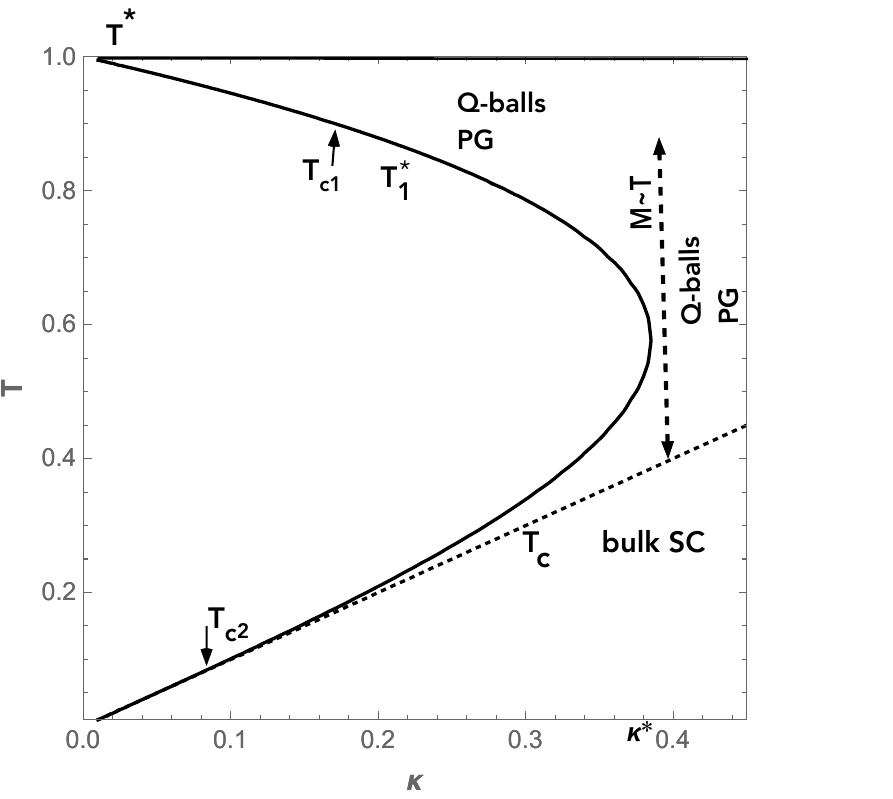}\vspace{0.1cm}
\caption{The phase diagram that follows from Eqs. (\ref{UFOIC}), (\ref{MSQRT}),  where $\kappa\equiv c \dfrac{4g\nu \varepsilon_0}{3}$, see text. The dashed line indicates interval of temperatures at coupling constant $\kappa^*=2\mu_0^3/(3\sqrt{3})$ in which linear temperature dependence $M\sim T$ is found from Eqs. (\ref{MSOLS}), (\ref{MTLIN}). Temperature is expressed in units of $T^*=\mu_0/2\pi$. The dotted line indicates superconducting transition temperature into bulk superconducting state (infinite Q-ball volume) found from solution of Eq. (\ref{UFOIC}). The contour plot $T^*_1$ as function of $\kappa$ is found from Eq. (\ref{MSQRT}). The arrows bars $1,2$ indicate points at which dynamic structure factor $S(Q_{CDW},\omega)\equiv I(\omega)$ is calculated from Eq. (\ref{MAPS}) and plotted below in Figs. \ref{0}, \ref{00}.}
\label{1}
\end{figure}
Namely, substituting Eq. (\ref{UFOA}) into the self-consistency equation (\ref{self}) one finds:
\begin{align}
& 0=(\mu _0 ^2 -\Omega^2)M^2+ gU_{f}(M)=(\mu _0 ^2 -\Omega^2)\left\{\left(M-\dfrac{g\nu \varepsilon_0}{6\sqrt{2}(\mu _0 ^2 -\Omega^2)}\right)^2-\right.\nonumber\\
&\left. -\left(\dfrac{g\nu \varepsilon_0}{6\sqrt{2}(\mu _0 ^2 -\Omega^2)}\right)^2\left(1-\Omega\dfrac{6\sqrt{2}(\mu _0 ^2 -\Omega^2)}{g\nu \varepsilon_0}\right)\right\}\,, 
  \label{selfi}
\end{align}
\noindent that trivially leads to the two-branches solution $M_{\pm}(\Omega)$ obtained previously (see Fig.3 in \cite{Mukhin(2022), Mukhin(2022_1)})from numerics:

\begin{align}
M_{\pm}= \dfrac{g\nu \varepsilon_0}{6\sqrt{2}(\mu _0 ^2 -\Omega^2)}\left(1\pm\sqrt{1-\Omega\dfrac{6\sqrt{2}(\mu _0 ^2 -\Omega^2)}{g\nu \varepsilon_0}}\right).
\label{MSOLS}
\end{align} 

\noindent First of all, introducing notation:

\begin{align}
\kappa=\dfrac{g\nu \varepsilon_0}{6\sqrt{2}}\equiv\dfrac{4g\nu \varepsilon_0}{3\cdot 8\sqrt{2}}\equiv c\dfrac{4g\nu \varepsilon_0}{3}\equiv\kappa\,;\;\; c=\dfrac{1}{8\sqrt{2}}\approx 0.09
\label{kappa}
\end{align}

\begin{figure}[H]
 \includegraphics[width=0.45\linewidth]{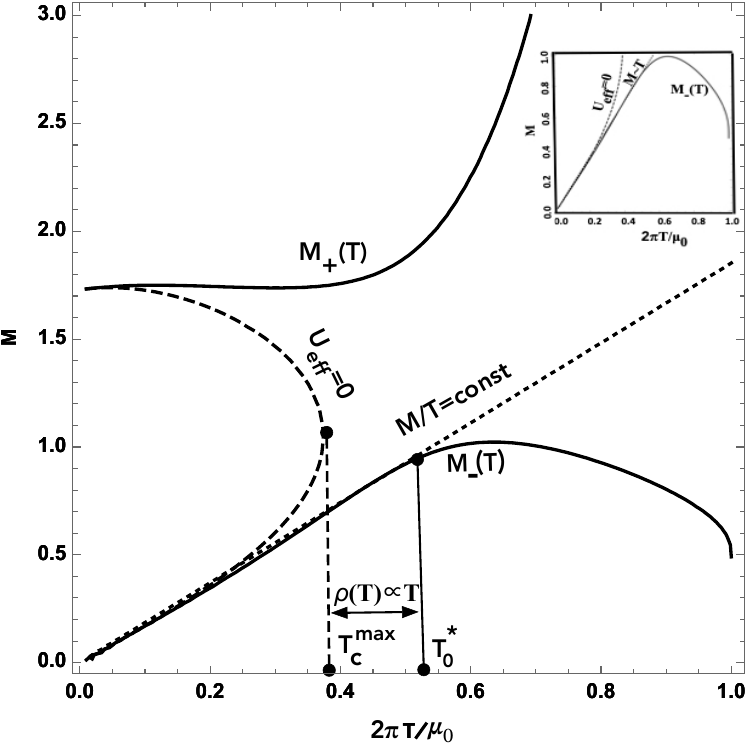}\vspace{0.1cm}
\caption{The numerical solutions of Eqs. (\ref{UFOIC}) and (\ref{selfi}) for the $M_{\pm}(T)$ branches (solid lines) and for the $U_{eff}(M)=0$ (long dashed line) at coupling constant $\kappa > \kappa^*=2\mu_0^3/(3\sqrt{3})$. The dotted line is guide for the eye demonstrating origin of the linear T-dependence of the Q-ball field amplitude $M$ in the interval of temperatures $T_c<T<{T_{0}}^* <T^*= \mu_0/2\pi$, as found analytically in Eq. (\ref{MTLIN}) from approximate expression for the $U_{f}$ in Eq. (\ref{UFOA}), see text. Temperature is expressed in units of $\mu_0/2\pi$ and amplitude $M$ in units of $\mu_0$. The inset reminds that only $M(T) < 1$ are allowed for the actual solutions due to the limitation for the amplitude of the field  $M(\tau,{\bf{r}})$ related with the local condensed spin/charge densities, see text after Eq. (\ref{Eu1})}
\label{MM}
\end{figure}
\noindent one finds the following equation that defines the boundary of the regions in the $\{\Omega,\kappa\}$ plane where both brunches $M_{\pm}$ in Eq. (\ref{MSOLS}) are real: 

\begin{align}
1-\Omega\dfrac{(\mu _0 ^2 -\Omega^2)}{\kappa}=0\,.
\label{MSQRT}
\end{align} 
\noindent Thus, one finds that Eq. (\ref{MSQRT}) coincides with condition previously obtained numerically \cite{Mukhin(2022), Mukhin(2022_1)} for the existence of  solution of the self-consistency equation (\ref{self1}), see Fig. \ref{1}.
\noindent 
In particular, a straightforward algebra gives from Eq. (\ref{MSQRT}) the result for the strength $\kappa^*=2\mu_0^3/(3\sqrt{3})$ corresponding to the touching point of the temperature boundaries of the strange metal and superconducting dome $T^*(\kappa^*)= T_c(\kappa^*)=\mu_0/(2\pi\sqrt{3})$. Here one notes that at $\kappa^*=2\mu_0^3/(3\sqrt{3})$ the bulk superconductivity transition temperature ${T^b}_c$ found from Eq. (\ref{UFOIC}) is below the Q-balls temperature $T_c(\kappa^*)$: ${T^b}_c(\kappa^*)=2\mu_0/(6\pi\sqrt{3}) < T_c(\kappa^*)=\mu_0/(2\pi\sqrt{3})$, which is obvious from the Fig. (\ref{UFOIC}), where bulk superconductivity temperatures are plotted with dotted line.  It is remarkable, that using neutron scattering results for $\mu_0$ related with effective superexchange coupling $J\approx 140 meV$ between neighbouring magnetic moments in high-T$_c$ cuprates \cite{Tranq0} one finds 'why  $T_c$ is high':  
\begin{align}
T_c(\kappa^*)=\mu_0/(2\pi\sqrt{3})\sim 100^\circ K  - 200^\circ K
\label{Tc}
\end{align} 
\noindent
Last, but not the least, it follows from Eq. (\ref{MSOLS}) that in the interval $\mu_0/(2\pi\sqrt{3})\leq T\leq\mu_0/(2\pi) $, i.e. in the temperatures interval $\{T_c(\kappa^*), T^*\}$ with  $T^*=\mu_0/(2\pi)$, the square root in Eq. (\ref{MSOLS}) could be expanded in the powers of  the second term, which is smaller then unity,
leading to the following expression for $M_{-}$ brunch:

\begin{align}
M_{-}\approx \dfrac{g\nu \varepsilon_0}{6\sqrt{2}(\mu _0 ^2 -\Omega^2)}\dfrac{\Omega 3\sqrt{2}(\mu _0 ^2 -\Omega^2)}{g\nu \varepsilon_0}\equiv \dfrac{\Omega}{2}=\pi T.
\label{MTLIN}
\end{align}
\noindent This remarkable result, meaning the linear temperature dependence of the Q-ball field amplitude $M$, leads to the linear temperature dependence of the electric resistivity due to scattering of electrons on the $M_{-}$ field Q-balls, as is shown in the next Section IV. 
The numerical results of finding $M(T)$ dependences after substitution into Eq. (\ref{UFOIC}) and Eq. (\ref{selfi}) of the complete expression for the effective potential energy $U_{f}$ from Eqs. (\ref{UFO1}), (\ref{UFOI1}) are plotted in Fig. \ref {MM}. 

\section{Electron scattering and resistivity of  Q-ball gas}
The Q-ball mechanism of the high-T$_c$ superconductivity and pseudo-gap phase in cuprates introduced previously \cite{Mukhin(2022), Mukhin(2022_1), Mukhin(2023)} is in essence a mechanism of  Cooper-pairing that occurs due to pairing of fermions via exchange with bosonic fluctuations of spin- or charge density waves (SDW/CDW) condensed locally into Q-balls, the nontopological solitons of thermodynamic quantum time crystals. The conserved Noether charge Q counts the total number of condensed bosonic fluctuations inside the Q-ball, and the basic internal rotation frequency of the Q-ball is bosonic Matsubara frequency $\Omega=2\pi T$ of the fundamental Fourier component of the SDW/CDW semiclassical fluctuation. The heterogeneous phase of Q-balls appears below T$^*$ temperature and exists down to the temperature T$_1^*$, that bounds from below the 'strange metal' phase. In the optimally doped case T$_1^*$ coincides with the top of the superconducting dome T$_c$ of the high-T$_c$ cuprates phase diagram  \cite{Mukhin(2022), Mukhin(2022_1)}. Below we demonstrate that influence of Q-balls on the electrical transport in the "strange metal" phase causes "Planckian"  \cite{Jan} linear temperature dependence of the normal metal resistivity \cite{Niven}. In short, since a Q-ball occupies finite space, there are outside electrons, that are not Cooper paired, and are scattered by the Q-ball SDW/CDW fluctuation. Demonstration of the fact that the T-linear temperature dependence of electrical resistivity of the "strange metal" phase  occurs due to electrons scattering on the Q-balls is the focus of the present work. Besides, dragged by electric field (unpinned) CDW Q-balls become sliding charge 'droplets', and hence, also contribute to the resistive normal current. This effect is considered below as well. Stability of Q-balls was proven for finite temperatures in \cite{Mukhin(2022), Mukhin(2022_1)} and long before that for the ground state of quantum matter \cite{Coleman (1985)}, \cite{Rosen}. 
  
\begin{figure}[H]
\includegraphics[width=0.45\linewidth]{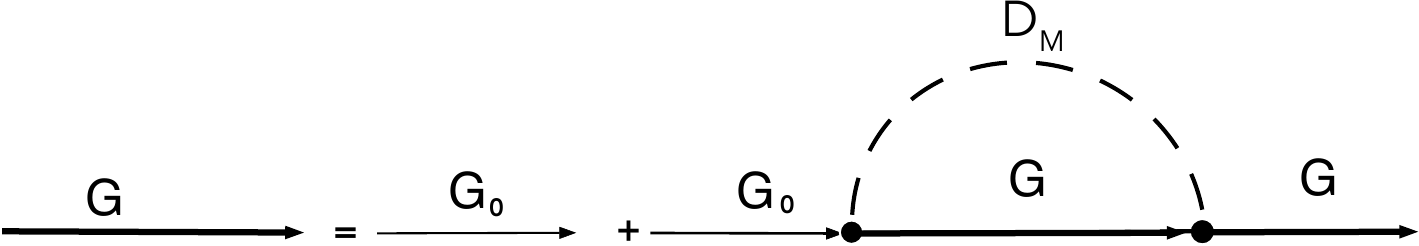}\hspace{0.2cm}
\caption{ The Dyson's equation for a fermion scattering by Q-balls of CDW/SDW bosonic field, see text. } 
\label{Dys2}
\end{figure}

To proceed one uses the Q-ball  - fermion interaction Hamiltonian in the form \cite{Mukhin(2022_1)}:
\begin{eqnarray}
\hat{H}_{int}=8\pi\kappa\sum_{\vec{p},\vec{q},i}e^{-i\vec{q}\vec{R}_i}\left(\dfrac{Mc^{+}_{{\bf{p}},\sigma}c_{{\bf{p-q}},\sigma}e^{-i\Omega(\tau+\tau_0)}}{(\kappa^2+(\vec{q}-\vec{Q})^2)^2}+ \dfrac{M^{*}c^{+}_{{\bf{p-q}},\sigma}c_{{\bf{p}},\sigma}e^{i\Omega(\tau+\tau_0)}}{(\kappa^2+(\vec{q}+\vec{Q})^2)^2}\right)
\label{intinst0}
\end{eqnarray}
\noindent where $\vec{Q}$ is either antiferromagnetic Brillouin zone  SDW nesting wave-vector, or CDW wave-vector connecting the hot spots of the Fermi surface, and $\kappa=1/R\propto {V}^{-1/3}$, and $R$, $V$, $M$ are Q-ball radius, volume and amplitude defined in Eqs. (\ref{step}), (\ref{Q})  and found self-consistently.
Summation over random coordinates $\vec{R}_i$ of the Q-ball centres is assumed in Eq. (\ref{intinst0}). The Dyson equation for the Green's function of electrons scattered on the Q-balls potential is presented in Fig. \ref{Dys2}. It follows from the well-known impurity scattering procedure \cite{AGD}, that averaging over the coordinates $\vec{R}_i$ of the Q-ball centres leads to the sum over double-scattered fermions on each Q-ball separately.

\noindent
In Fig. \ref{Dys2} the heavy and thin lines are fermionic temperature Green's functions  $G({r}-{r}', \tau-\tau')$ and $G_0({r}-{r}', \tau-\tau')$ respectively, that depend on the differences of the $D+1$ coordinates after averaging over positions of the Q-balls in space and Matsubara time origin $\tau_0$. Dots are vertices of fermion- Q-ball field $M$ interaction introduced in Eq. (\ref{intinst0}).
The M-field bosonic Green's function ${D}_M$, that follows from Eq. (\ref{intinst0}) after averaging over positions of the centres of  the Q-balls $\vec{R}_i$ and Matsubara time zero-origin $\tau_0$ is:
\begin{eqnarray}
{D}_M(\vec{q},\omega)=\left(8\pi M\kappa\right)^2\left\{\dfrac{\delta_{\omega,\Omega}}{(\kappa^2+(\vec{q}-\vec{Q})^2)^4}+ \dfrac{\delta_{\omega,-\Omega}}{(\kappa^2+(\vec{q}+\vec{Q})^2)^4}\right\}
\label{DMq}
\end{eqnarray}
\noindent It is remarkable that due to semiclassical nature of the Q-ball fluctuation its Green's function ${D}_M(\vec{q},\omega)$ possesses only single frequency $\Omega$ which is self-consistently determined from Eqs. (\ref{self}) and (\ref{UFO1}), (\ref{UFOI1}). Then, taking the Green's function of the scattered fermions in the form: 
\begin{eqnarray}
G(\vec{p},\omega)=\dfrac{1}{i\omega-\xi(\vec{p})-\bar{G}(\vec{p},\omega)};\quad \xi(\vec{p})=\varepsilon(\vec{p})-\mu\,,
\label{G}
\end{eqnarray}
\noindent where $\mu$ is the chemical potential, and using the Dyson's equation in Fig. \ref{Dys2}, one finds the following equation for the self-energy function 
$\bar{G}$: 

\begin{eqnarray}
\bar{G}(\vec{p},\omega)=\sum_Q \bar{n}_{Q}M^2\dfrac{(8\pi\kappa)^2}{(2\pi)^3}\int d^3\vec{q}\quad \dfrac{G(\vec{p}-\vec{q},\omega-\Omega)+G(\vec{p}+\vec{q},\omega+\Omega)}{(\kappa^2+(\vec{q}-\vec{Q})^2)^4}\,.
\label{DGn}
\end{eqnarray}
 \noindent where $\bar{n}_{Q}$ is density of Q-balls  with "charge" Q defined as:
 \begin{eqnarray}
&& \bar{n}_{Q}=\dfrac{1}{V}G_{Q}C\exp{-\dfrac{E_{Q}}{k_BT}}=\dfrac{C}{V_{Q}}\exp{-\dfrac{2Q\Omega}{gk_BT}}=\dfrac{4\pi}{gV_{Q}}\exp{-\dfrac{4\pi Q}{g}}\,,\nonumber\\
&& G_{Q}=\dfrac{V}{V_{Q}},\quad C=\dfrac{4\pi}{g}\,,
 \label{LL5}
 \end{eqnarray}  
 \noindent
  Below it is assumed for simplicity that major scattering involves the fermions that occupy hole pockets in the Brillouin zone of doped cuprates with $\vec{Q}=\vec{Q}_{SDW}$ being approximately magnetic Brillouin zone wave vector \cite{Mukhin(2022), Mukhin(2022_1)}, or $\vec{Q}=\vec{Q}_{CDW}$ and connects hot spots on the Fermi surface. Therefore, it is assumed that both $\vec{Q}$-vectors connect quasiparticle states of the opposite energies with respect to Fermi level, i.e. quasi-holes with quasi-electrons and vice versa : $\xi(\vec{p}\pm \vec{Q})= -\xi(\vec{p})$. Hence, using the latter equalities it is straightforward to change integration vector in the integral equation (\ref{DGn}): $\vec{q}-\vec{Q}\rightarrow \vec{q}$, that leads to:
 
 \begin{eqnarray}
\bar{G}(\vec{p},\omega)=\sum_Q \bar{n}_{Q}M^2\dfrac{(8\pi\kappa)^2}{(2\pi)^3}\int d^3\vec{q}\quad \dfrac{G(\vec{p}-\vec{q},\omega-\Omega)+G(\vec{p}+\vec{q},\omega+\Omega)}{(\kappa^2+{q}^2)^4}\,.
\label{DGnp}
\end{eqnarray}
\noindent Then, assuming: $\xi=\varepsilon(\vec{p})-\mu=p^2/2m-\mu$, and changing  the integration variables (compare \cite{AGD}):

 \begin{eqnarray}
\int d^3\vec{q}=\frac{2\pi m}{p}\int^{\infty}_{0} qdq\int^{\xi_{+}}_{\xi{-}}d\xi\,;\;\xi_{\pm}=\xi(p\pm q)\,,
\label{chan}
\end{eqnarray}
\noindent
 one rewrites Eq. (\ref{DGnp}) in the form:
 \begin{eqnarray}
\bar{G}(\vec{p},\omega)=&&\sum_Q \bar{n}_{Q}M^2\dfrac{(8\pi\kappa)^2}{(2\pi)^2}\frac{m}{p}\int^{\infty}_{0} qdq\int^{\xi_{+}}_{\xi{-}}d\xi\quad \dfrac{1}{(\kappa^2+{q}^2)^4}\left\{\dfrac{1}{i(\omega-\Omega)+\xi-\bar{G}_{-}}+\right.\nonumber\\
&&\left.\dfrac{1}{i(\omega+\Omega)+\xi-\bar{G}_{+}}\right\}\,;\quad \bar{G}_{\mp}=\bar{G}(\xi,\omega\mp\Omega)
\label{SEG}
\end{eqnarray}
\noindent Now, allowing for the relation justified aposteriori: $\bar{G}_{\mp}=\bar{G}$,  Eq. (\ref{SEG}) reads:

 \begin{eqnarray}
&&\bar{G}=\sum_Q \bar{n}_{Q}M^2\dfrac{(8\pi\kappa)^2}{(2\pi)^2}\frac{m}{p}\int^{\infty}_{0} \dfrac{qdq}{(\kappa^2+{q}^2)^4}\int^{\xi_{+}}_{\xi{-}}d\xi \dfrac{2(i\omega+\xi-\bar{G})}{(i\omega+\xi-\bar{G})^2+\Omega^2}
\label{SEG1}
\end{eqnarray}
\noindent Next, analytic continuation of Eq. (\ref{SEG1}) to the  real axis of frequencies, $i\omega\rightarrow \omega$, gives:
 \begin{eqnarray}
&&\bar{G}(\vec{p},\omega)=\sum_Q \bar{n}_{Q}M^2\dfrac{(8\pi\kappa)^2}{(2\pi)^2}\frac{m}{p}\int^{\infty}_{0} \dfrac{qdq}{(\kappa^2+{q}^2)^4}\ln{
\dfrac{(\omega+\xi_{+}-\bar{G})^2+\Omega^2}{(\omega+\xi_{-}-\bar{G})^2+\Omega^2}}
\label{DGR}
\end{eqnarray}
\noindent The Q-ball form factor $\propto (\kappa^2+{q}^2)^{-4}$ reduces integration over $q$ to the interval $0\leq q\leq \kappa$ and, therefore, allowing for the mesoscopic Q-ball sizes \cite{Mukhin(2023)}: $R_Q^{-1}\sim \kappa\ll p$,  it is fare to approximate the above relation expanding $\xi_{\pm}$ to the first order in $q\sim\kappa$: 
\begin{eqnarray}
\xi_{\pm}=\xi(p\pm q)\approx \xi(p)\pm vq\;;\quad v\equiv \dfrac{\partial \xi(p)}{\partial p}
\label{xi}
\end{eqnarray}
\noindent Hence, one finds from Eq. (\ref{DGR}) the following 'on shell',  $\omega+\xi(p)=0$, equation for $\bar{G}(\vec{p},\omega)$:
\begin{eqnarray}
\bar{G}(\vec{p},\omega)=\sum_Q \bar{n}_{Q}M^2\dfrac{(8\pi\kappa)^2}{(2\pi)^2v}\int^{\infty}_{0} \dfrac{qdq}{(\kappa^2+{q}^2)^4}\ln{
\dfrac{\Omega^2- 2vq\bar{G}}{\Omega^2+2vq\bar{G}}}
\label{G_R}
\end{eqnarray}
\noindent Now, assuming $\bar{G}$ to be responsible for electrons damping rate and thus purely imaginary, one finally finds after integration in (\ref{G_R}) an equation for $\bar{G}$:

\begin{eqnarray}
\bar{G}=-\sum_Q \bar{n}_{Q}M^2\dfrac{4\pi}{2\Omega^2\kappa^3}{\left[\bar{G}+\dfrac{4v^2\kappa^2\bar{G}^3}{3\Omega^4}\right]}\equiv - \left[I_1\bar{G}+I_2\bar{G}^3\right]
\label{Gfin}
\end{eqnarray}
\noindent Using now definition for $\bar{n}_{Q}$ from Eq. (\ref{LL5}) and relation $\kappa =1/R_Q$, where $R_Q$ is Q-ball radius, substituting summation over $Q$ by integration, expressing $V_Q$ via $M$ and $Q$ using Eq. (\ref{Q}), and allowing for the scaling of the Q-ball amplitude with temperature in Eqs. (\ref{Mstar}), (\ref{MTLIN}): $M=s\Omega$, $s>1$, one finds:

\begin{eqnarray}
I_1=\sum_Q \bar{n}_{Q}M^2\dfrac{4\pi}{2\Omega^2\kappa^3}=\int_0^{\infty}\dfrac{6M^2P(Q)dQ}{\Omega^2 4}=\dfrac{3M^2}{2\Omega^2}=\dfrac{3s^2}{2}\,;\quad \dfrac{1}{\kappa^3}=\dfrac{3V_Q}{4\pi}
\label{IDN1 }
\end{eqnarray}
\noindent 
The coefficient $I_2$ in front of $\bar{G}^3$ in Eq.(\ref{Gfin})  is more elaborate: 

\begin{eqnarray}
I_2=\sum_Q \bar{n}_{Q}M^2\dfrac{4\pi}{2\Omega^2\kappa^3}\dfrac{4v^2\kappa^2}{3\Omega^4}=4\int_0^{\infty}\dfrac{M^2P(Q)v^2dQ}{2\Omega^6 V_Q^{\frac{2}{3}}}
\left(\dfrac{4\pi}{3}\right)^{\frac{2}{3}}\,;\quad {\kappa^2}=\left(\dfrac{4\pi}{3V_Q}\right)^{\frac{2}{3}}
\label{IDN2}
\end{eqnarray}
\noindent Hence, 
 \begin{eqnarray}
I_2=4\left(\dfrac{4\pi}{3}\right)^{\frac{2}{3}}\dfrac{v^2s^{\frac{10}{3}}}{2\Omega^2}\int_0^{\infty}\dfrac{P(Q)dQ}{Q^{\frac{2}{3}}}=\dfrac{\tilde{C}}{\Omega^2}\,; 
\tilde{C}\equiv4\dfrac{(4\pi)^{\frac{4}{3}}v^2s^{\frac{10}{3}}}{2(3g)^{\frac{2}{3}}}\int_0^{\infty}\dfrac{e^{-x}dx}{x^{\frac{2}{3}}}
\label{IDN21}
\end{eqnarray}                                                          
\noindent
Solving Eq. (\ref{Gfin}) with the aid of relations (\ref{IDN1 }) and (\ref{IDN21}) one finds the following relation for the fermionic quasiparticle lifetime due to Q-ball scattering ($\pm$ sign below is chosen depending on retarded- or advanced Green's function is considered), $\tau_Q$: 
\begin{eqnarray}
\bar{G}=\pm\frac{i}{\tau_Q}\,;\quad \dfrac{1}{\tau_Q}= \sqrt{\dfrac{1+I_1}{I_2}}= \dfrac{\Omega}{\sqrt{\tilde{C}}}\sqrt{1+{3s^2}/{2}}\propto T\,.
 \label{DGtau}
\end{eqnarray}  
\noindent The above result is remarkable, since it demonstrates that linear temperature dependence of the fermionic inverse lifetime arises due to Q-ball scattering in the whole temperature interval $T_1^*<T<T_0^*$, thus providing origin of the "strange metal" behaviour.  The bosonic frequency $\Omega=2\pi T$ of the quantum thermodynamic Q-ball time crystal plays the role of  a scattering rate $1/\tau\propto\Omega$ for the fermions in the Q-ball semiclassical field, manifesting the prominent 'Planckian' scattering rate behaviour \cite{Jan}. It follows also from Eq. (\ref{DMq}), that $D(\pm\vec{q})$ plays the role of  $\pm \Omega$ Fourier components of the Q-ball field propagator modulo Q-ball density $\bar{n}_Q$. Simultaneously, the CDW/SDW wave vector $\vec{Q}$ entering propagator $D(\vec{q})$, causes anisotropy of the scattering rate, thus explaining  'quantum nematic' behaviour known for high-$T_{c}$ cuprates \cite{Hinkov}:

\begin{eqnarray}
\sigma_{i,j}\propto \dfrac{Q_iQ_j \tau_Q}{\vec{Q}^2}\,,
 \label{nem}
\end{eqnarray}  
\noindent where $\sigma_{i,j}$ is electron conductivity tensor.

\section{Electron resistivity due to Q-balls slide}
The picture of 'free' fermions scattered by a gas of  randomly distributed in space Q-balls considered in the previous Section might be not complete in the case of Q-balls, that are not pinned to the lattice. Namely, one may consider contribution to the electrical resistivity coming from the slide of the Q-ball CDW as a whole in a weak electric field. To calculate this contribution one may use a method described in \cite{Alex} by adding potential energy term of a Q-ball charge density in a homogeneous constant electric field: $\phi =- e\vec{r}\vec{E}$, thus adding an extra term to the Euclidean action in Eq. (\ref{Eu}) and, correspondingly, modifying the saddle-point equation, that becomes then:
 
\begin{eqnarray}
&&\dfrac{\delta S_{M}}{\delta M^*(\tau,{\bf{r}})}=-\partial^2_\tau M(\tau,{\bf{r}})-s^2\sum_{\alpha={\bf{r}}}\partial^2_\alpha M(\tau,{\bf{r}})+\mu _0 ^2 {M(\tau,{\bf{r}})}+gM(\tau,{\bf{r}})\dfrac{\partial U_f}{\partial 
|M(\tau,{\bf{r}})|^2 }\nonumber \\
&&- 2i\Omega(\partial_{\tau}+\dfrac{ie\phi}{\hbar})M(\tau,{\bf{r}})=0\label{L1}
\end{eqnarray} 
\noindent Solving this equation expressed via Fourier transformed function $M(\Omega, \vec{p})$ to the first order in potential $\phi$ Fourier component, one finds:   

\begin{eqnarray}
M=M_0+M_1;\; M_1(\pm \Omega,\vec{p})=\dfrac{2\Omega e\phi  M_0(\pm \Omega,\vec{p})}{\hbar(\mu_0^2-\Omega^2)}=\dfrac{2\Omega e\vec{E}}{\hbar(\mu_0^2-\Omega^2)}\dfrac{i\partial M_0(\pm \Omega,\vec{p})}{\partial\vec{p}}      \label{M1}
\end{eqnarray}
\noindent where $M_0$ reads:

\begin{eqnarray}
M_0(\pm \Omega, \vec{p})= \dfrac{8\pi\kappa M}{(\kappa^2+(\vec{p}\mp \vec{Q})^2)^2 }
     \label{M0}
\end{eqnarray}
\noindent Then, to the first order in electric field $\vec{E}$ the Q-ball sliding CDW current density reads:

\begin{eqnarray}
&&\vec{j}=-\dfrac{ie\hbar}{4m}\sum_{q}(M^{*}\vec{\nabla}M-M\vec{\nabla}M^{*})=\dfrac{e^2}{2m}\dfrac{\Omega}{(\mu_0^2-\Omega^2)}\sum_{p}\vec{p}\vec{E}\cdot\dfrac{\partial}{\partial\vec{p}} \left[M_0(\Omega,\vec{p})^2+\right.\nonumber\\
&&\left. M_0(-\Omega,\vec{p})^2\right]\equiv \vec{E}\sigma_{CDW}
     \label{jCDW}
\end{eqnarray}
\noindent and hence: 
\begin{eqnarray}
\sigma_{CDW}\propto\dfrac{e^2\Omega M^2}{m(\mu_0^2-\Omega^2)\kappa^3}
 \label{sCDW}
\end{eqnarray}
\noindent First, expression in Eq.(\ref{sCDW}) is remarkably different from expression for the electrical conductivity due to scattering of the 'free electrons' on the Q-balls. Namely, the pronounced nematicity of the conductivity tensor in Eq. (\ref{nem}) is manifestly absent in Eq. (\ref{sCDW}). This points to a hydrodynamic character of the Q-ball CDW slide in external electric field. Next, it is instructive to apply above result to the vicinity of T$^*$ temperature, since the power indices for the temperature dependencies of the Q-ball parameters where found earlier \cite{Mukhin(2022), Mukhin(2022_1)}. Taking into account that superconducting gap approaches zero at T$^*$ according to Eq. (\ref{g0star}), and following Ginzburg-Landau theory of the superconducting order parameter, see e.g. Chapt.17  \cite{Alex}, one finds that minimal radius $R_{min}$ of the Q-ball with superconducting condensate inside diverges as \cite{Mukhin(2022_1)}: 

\begin{eqnarray}
R_{min}\propto \dfrac{1}{g_0}\sim \dfrac{1}{(T^*-T)^{\frac{1}{5}}}\,.
 \label{Rmin}
\end{eqnarray}
\noindent
Then, using definition of T$^*$ in Eq. (\ref{Mstar}), one rewrites expression Eq. (\ref{sCDW}) in the asymptotic form:

\begin{eqnarray}
\sigma_{CDW}\propto\dfrac{e^2\Omega M^2}{m(\mu_0^2-\Omega^2)\kappa^3}\sim \dfrac{{R_{min}}^3}{T^*-T}\propto\dfrac{1}{(T^*-T)^{{8}/{5}}}\equiv \dfrac{1}{(T^*-T)^{1.6}}
 \label{sT}
\end{eqnarray}
\noindent This critical behaviour significantly differs from Ginzburg-Landau theory prediction for the 3D case in the vicinity of superconducting transition temperature T$_c$ \cite{Alex}:
\begin{eqnarray}
\sigma_{GL}\propto\dfrac{1}{(T-T_c)^{\gamma}};\; \gamma={1}/{2}
\label{sGL}
\end{eqnarray}
\noindent and is most close to the 1D case, $\gamma=3/2$, \cite{Alex}.

\section{Diamagnetic response of Q-ball gas}

\begin{figure}
\centering
    \includegraphics[width=0.8\linewidth]{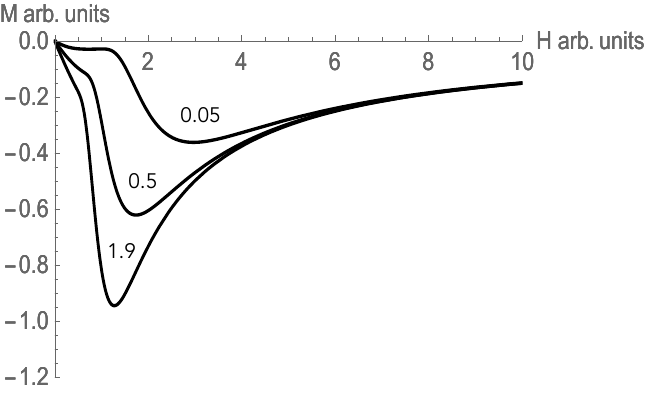}
\caption{ Density of diamagnetic moment of the Q-balls gas in the PG phase $T_{1}^*(\kappa)<T<T^*$, curves 1-3 correspond to different values of departure of the temperature $T$ from $T^*$:  $T^*-T \equiv \mu_0/(2\pi )-T$, indicated in arb. units, see Fig. \ref{1} and Eqs. (\ref{Mstar}), (\ref{M1}), (\ref{M2}).}
\label{dia}
\end{figure}

It is straightforward to apply presented above picture of Q-ball gas in high-T$_c$ superconductors for description of experimentally discovered diamagnetic behaviour above T$_c$ in cuprates \cite{Li, Zaanen}. 
Again, as in Eq. (\ref{LL5}) using the concept of the phase space of the Q-balls formed by the values of the 'Noether charge' $Q$ and discrete
 values of the Matsubara frequencies $\Omega_n\equiv 2\pi nT$, $n=1,2,...$, and counting the number of the different 'positions' of a Q-ball in the real space as  ${V}/{V_{Q,n}}$, where $V$ is the volume of the system and the Q-ball volume is determined using the 'charge' Q conservation law Eq. (\ref{Q}):

 \begin{eqnarray}
V_{Q,n}\equiv \dfrac{4\pi R^3}{3}= \dfrac{Q}{\Omega_n M^2}\; , \label{QQ}
\end{eqnarray}
\noindent one finds the following expression for the partition function of the Q-balls gas in the temperature range where it exists, $T_{1}^*(\kappa)<T<\mu_0/(2\pi )$, see Fig. \ref{1}:

\begin{eqnarray}
Z_Q=\sum_{Q,n}\frac{1}{N!}\left[\int_{Q_m}^{Q_H}dQ\frac{V}{V_{Q,n}}\exp\left\{-\left[\dfrac{2Q\Omega_n}{gT}-\dfrac{M_QH}{T}\right]\right\}\right]^N   \;,
  \label{MQ}
\end{eqnarray}  

\noindent
 The Q-ball energy in the first term of the Boltzmann's expression in the brackets in Eq. (\ref{MQ}), $E_Q/T$, is taken from the self-consistency Eq. (\ref{aQ}). The lower and upper bounds in the integral over $dQ$ are as follows. The smallest value of $Q=Q_m$ is obtained from Eq. (\ref{QQ}) for the Q-ball of the size $R_m$ bound from below by the Landau correlation length $\xi$, see Eq. (\ref{Rmin}): 

\begin{eqnarray}
&{Q_m}={\Omega M^2}\dfrac{4\pi R_m^3}{3}\,,\;R_m=\xi\equiv \pi\sqrt{\dfrac{\hbar^2}{4mbg_0^2}}\;. \label{Qmm}
\end{eqnarray}
\noindent with $g_0$ defined by Eq. (\ref{g0star}). The upper bound $Q_H$ in the integral in Eq. (\ref{MQ}) is obtained as follows:

\begin{eqnarray}
&{Q_H}={\Omega M^2}\dfrac{4\pi R_H^3}{3}\,,\;R_H=\dfrac{\delta_LH_c\sqrt{20}}{H}\,,\;\delta_L=\dfrac{\sqrt{mc^2}}{\sqrt{4\pi n_se^2}}\,, \label{QH}
\end{eqnarray}
\noindent where $R_H\ll \delta_L$ is the maximum radius of a small superconducting sphere \cite{LL9}, at which it remains  superconducting in magnetic field $H$, and $\delta_L$ is London penetration depth, $H_c$ is critical magnetic field of the bulk superconductor material,  $n_s\approx 2\pi \nu T_c/3$ is superconducting electrons density, as derived in \cite{Mukhin(2022)} in accord with Uemura plot behaviour \cite{Uemura}, with $2\pi T_c\equiv\Omega_c$ given in Eq. (\ref{Mg00S}), $\nu$ is the bare fermionic density at the Fermi level, $m$ is electron mass, and $c$ is light velocity. 
\noindent The next term, $-M_QH/T$, in the Boltzmann's expression in the brackets in Eq. (\ref{MQ}) is the energy of diamagnetic moment $M_Q$ in magnetic field $H$:

\begin{eqnarray}
&{M_Q}=-\dfrac{R^5H}{30\delta_L^2}H=-\left(\dfrac{3Q}{4\pi M^2\Omega}\right)^{\frac{5}{3}}\dfrac{H^2}{30\delta_L^2}\,, \label{MQLL}
\end{eqnarray}
\noindent where $M_Q$ is projection of diamagnetic moment of  a Q-ball on the magnetic field direction $\vec{H}$. The Q-ball is regarded as a small superconducting sphere of radius $R\ll \delta_L$ possessing diamagnetic moment in magnetic field $H$ \cite{LL9}. In the last equality in Eq. (\ref{MQLL}) $R$ is  substituted via the expression $R=R(Q)$ obtained from the Q-ball 'charge' $Q$ conservation relation Eqs. (\ref{Q}), (\ref{QQ}). Composing altogether the above relations one finds the following expression for the free energy of the Q-ball gas:

\begin{eqnarray}
&F=-T\ln{Z_Q}\,,\; Z_Q=\displaystyle \sum_{n,N}\dfrac{G_n^N}{N!}\equiv \exp{G_n}\,,\label{FQ}\\
&G_n= \displaystyle\int_{Q_m}^{Q_H}dQ\dfrac{V\Omega_n M^2}{Q}\exp\left\{-\left[\dfrac{2Q\Omega_n}{gT}+\left(\dfrac{3Q}{4\pi M^2\Omega_n}\right)^{\frac{5}{3}}\dfrac{H^2}{30\delta_L^2T}\right]\right\}   \;,\label{GQ}\\
&Q_H=\dfrac{\delta_L^3H_c^3}{H^3}\dfrac{4\pi\Omega_nM^2 20^{\frac{3}{2}}}{3}  \label{QHF}
\end{eqnarray} 
\noindent In the highest temperature interval $T_{1}^*(\kappa)<T<\mu_0/(2\pi n)$ one takes integer $n=1$, see Eq. (\ref{MSQRT}) and Fig. \ref{1}, and then for the free energy of the Q-balls gas and density of its diamagnetic moment $<M_Q>/V$ one finds: 

\begin{eqnarray}
&F=-TG_{n=1}\equiv -TG\,,\; <M_Q/V>=T\dfrac{\partial G}{V\partial H}\equiv- M_1-M_2\,,\label{Fn1}\\
&M_1=\dfrac{2H3^{5/3}}{30\delta_L^2(4\pi)^{5/3}( M^2\Omega)^{2/3}}\displaystyle\int_{Q_m}^{Q_H}dQ{Q}^{2/3}\exp\left\{-\left[\dfrac{2Q\Omega}{gT}+\left(\dfrac{3Q}{4\pi M^2\Omega}\right)^{\frac{5}{3}}\dfrac{H^2}{30\delta_L^2T}\right]\right\}\,,\label{M1}\\
&M_2=\dfrac{3\Omega M^2}{H}\exp\left\{-\left[\dfrac{2Q_H\Omega}{gT}+\left(\dfrac{3Q_H}{4\pi M^2\Omega}\right)^{\frac{5}{3}}\dfrac{H^2}{30\delta_L^2T}\right]\right\}\,,\label{M2}
\end{eqnarray}
\noindent where one has to substitute solution $M=M(\Omega)$ of the self-consistency Eq. (\ref{aQ}) using e.g. solutions from Eq. (\ref{Mstar}), or in the form of the two-branches solution Eq. (\ref{MSOLS}). This leads to the following dependence found numerically from Eqs. (\ref{M1}), (\ref{M2}) above, see Fig. \ref{dia}.

\section{"Hourglass" magnetic spectrum and splitting of the in-plane phonon brunches into the softened and hardened ones due to Q-balls scattering}

\begin{figure}
\centering
    \includegraphics[width=0.8\linewidth]{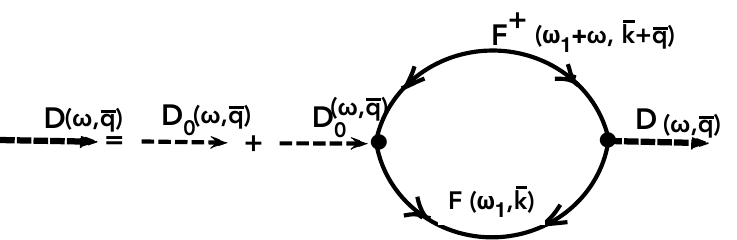}
\caption{ The Feynman diagram of the Dyson equation for the Green's function $D(\omega,\vec{q})$ of magnons/phonons with the self-energy (closed fermionic loop of Gor'kov's anomalous Green's functions $F$, $F^+$) caused by the presence of superconducting condensates inside the  Q-balls of coherently condensed antiferromagnetic/charge  density fluctuations with SDW/CDW wave vectors $\vec{Q}=\vec{Q}_{SDW}$ and $\vec{Q}=\vec{Q}_{CDW}$ respectfully.}
\label{101}
\end{figure}
\noindent
The scattering of  quantum spin excitations and phonons on the condensates of Cooper/local pairs inside the Q-balls induces self-energy, see Fig.\ref{101}, that consists of  closed fermionic loop of Gor'kov's anomalous Green's functions $F$, $F^+$, the latter are found in Eq. (\ref{Fam}) above.
The difference of the two cases of  the Q-balls formed by SDW and CDW fluctuations causes different patterns of the magnons and phonons dispersions in the vicinities of the $Q_{SDW}$ and $Q_{CDW}$ wave vectors in the brillouin zone. We start from the  $Q_{SDW}$ case.

\subsection {Hourglass magnetic spectrum  due to SDW Q-balls scattering} 

The Dyson equation for the Green's function $D(\omega, \vec{q})$ of magnons/phonons, that will be defined below by a particular choice of the vicinity of the wave-vector $\vec{q}$ and corresponding bare dispersion $\omega_0(\vec{q})$, takes the form \cite{AGD}:
\begin{eqnarray}
&D(\omega,\vec{q})=D_{0}(\omega,\vec{q})\left(1-D(\omega,\vec{q})\Pi_{F}(\omega,\vec{q})\right)\,,\label{DD}\\
&D_{0}(\omega,\vec{q})=-\dfrac{{\omega_{0}}^{2}(\vec{q})}{\omega^{2}+{\omega_{0}}^{2}(\vec{q})}\,,\omega=2\pi n\,;\label{D0}\\
&\Pi_{F}(\omega,\vec{q})={g^{2}_{bf}}T\sum_{\vec{k},\omega_{1}}\overline{F}_{\sigma,\bar{\sigma}}(\omega_{1},{\vec{k}}){F}_{\bar{\sigma},\sigma}(\omega_{1}+\omega,{\vec{k}}+{\vec{q}})\,,\label{Pif}
\end{eqnarray} 
\noindent where Matsubara bosonic frequency $\omega$ is defined by any integer $n$ and coupling constant $g_{bf}$ describes coupling of spin/phonon excitations to the fermions. Here one important observation is in order. 
\begin{figure}
\centering
    \includegraphics[width=0.8\linewidth]{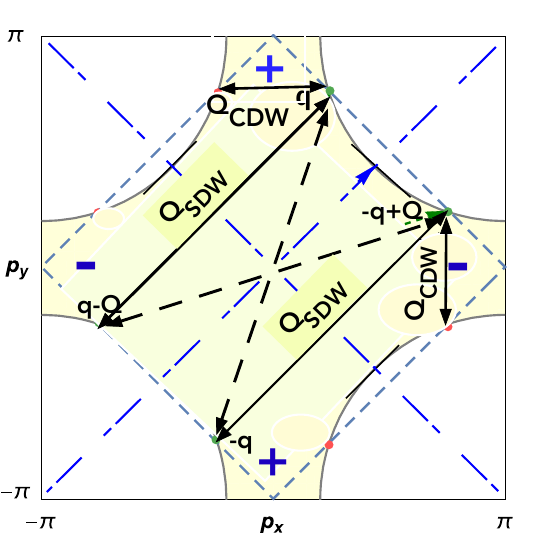}
\caption{'Nesting' wave-vectors  $\vec{Q}_{SDW}, \vec{Q}_{CDW}$ in the Brillouin zone of high-T$_c$ cuprates bare Fermi surface. The $\pm$ signs mark different sectors of the superconducting order parameter in the case of d-wave pairing symmetry (divided by dashed lines) related with spin-waves (magnons) collective antiferromagnetic fluctuations of the SDW type. The CDW type collective fluctuations wave vectors connect 'nested'/hot-spot regions of the Fermi surface and correspond to s-wave pairing of the superconducting order parameter. See text for explanations.}
\label{QF}
\end{figure}
\noindent
In the case of  the coupling with magnons and the Green's function with momenta close to the anti-ferromagnetic wave-vector $\vec{Q}_{SDW}$ corresponding to SDW fluctuations, one considers $D_{\sigma,\bar{\sigma}}(\omega,\vec{q}) $ component of the magnons Green's function in the Dyson's equation (\ref{DD}) above and correspondently the polarisation loop in Eq. (\ref{Pif}) contains then pre-factor $\sigma\bar{\sigma}=-1$, which is compensated by discussed previously, see end of Section 3 \cite{Mukhin(2022_1)}, d-wave symmetry of the superconducting order parameter known for cuprates \cite{campuzano}: ${F}_{\sigma,\bar{\sigma}}(\omega,{\bf{k}})= -{F}_{\sigma,\bar{\sigma}}(\omega,{\vec{k}}-{\vec{Q}}_{SDW})$.  On the other hand, when coupling with phonons is considered close to the 'nesting' wave-vector $\vec{Q}_{CDW}$ in the Brillouin zone corresponding to the CDW fluctuations, then the pre-factor $\sigma\bar{\sigma}$ is missing in the polarisation loop in Eq. (\ref{Pif}), and corresponding superconducting order parameter is of s-wave symmetry \cite{Annette}: ${F}_{\sigma,\bar{\sigma}}(\omega,{\vec{k}})= {F}_{\sigma,\bar{\sigma}}(\omega,{\vec{k}}-{\vec{Q}}_{CDW})$, see Fig. \ref{QF}. Hence, the expressions for the polarisation loop and Green's function of magnons with the wave vectors close to the antiferromagnetic wave-vector $Q_{AF}=\left\{\pi,\pi\right\}$ in the inverse crystal lattice units take the following form obtained with the use of Eq. (\ref{Fam}):

\begin{eqnarray}
D(\omega,\vec{q})=-\dfrac{\omega^{2}_{0}(\vec{q})}{\omega^{2}+{\omega^{2}_{0}}(\vec{q})\left(1-\Pi_{F}(\omega,\vec{q})\right)}\,,\omega=2\pi n\,;\label{DC}
\end{eqnarray}
\begin{eqnarray}
&\Pi_{F}(\omega,\vec{q})=g^{2}_{bf}\dfrac{\tanh{\left(\dfrac{g_0}{2T}\right)}}{g_0(4g_0^2+\omega^2)}\int \dfrac{d^3\vec{k}}{(2\pi\hbar)^{3}}\sqrt{{g_0}^2-\varepsilon^{2}(\vec{k})}\sqrt{{g_0}^2-\varepsilon^{2}(\vec{k}+\vec{q})}\equiv\nonumber\\
& \equiv g^{2}_{bf}\dfrac{\tanh{\left(\dfrac{g_0}{2T}\right)}}{g_0(4g_0^2+\omega^2)}\cdot I_{\vec{q}}\label{Pci}
\end{eqnarray} 
\noindent 
\begin{figure}
\centering
    \includegraphics[width=0.8\linewidth]{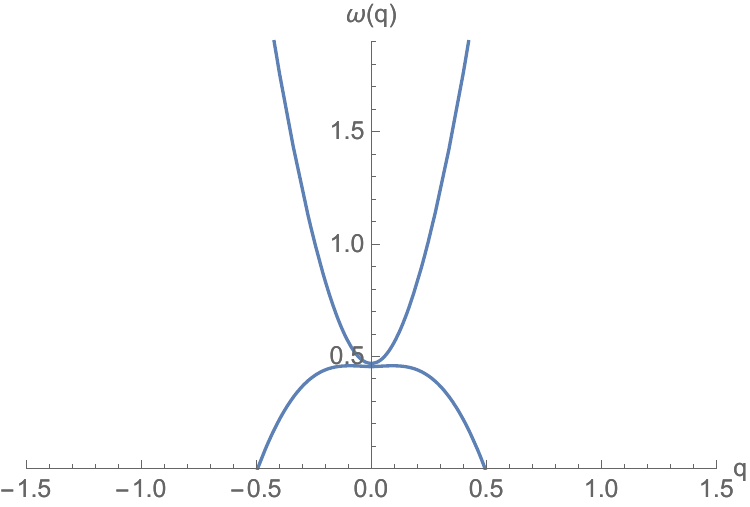}
\caption{Calculated hourglass dispersion $\omega({q})$ of magnetic excitations in the vicinity ${q}=|\vec{p}-\vec{Q}_{AFM}|$ of the wave vector $\vec{Q}_{AFM}$ of  coherently condensed antiferromagnetic spin density fluctuations forming the  Q-balls. }
\label{MMM}
\end{figure}
\noindent
Here we first take the bare fermionic dispersion $\varepsilon(\vec{k})$ along the 'nesting' vector $\vec{Q}_{SDW}$ close to the bare Fermi surface at the extended Van Hove singularity for a one-dimensional dispersion function: $\varepsilon(\vec{k})=v_f (k-p_F)$ and 'nesting' condition along the $\vec{Q}_{SDW}$ would be then: $\varepsilon(\vec{k}-\vec{Q}_{SDW})= - v_f (k-2p_F+p_F)\equiv -v_f (k-p_F)$  with the corresponding $2p_F$ : ${Q}_{SDW}=2p_F$, see Fig. \ref{QF}. Hence, calculation of the polarisation loop $\Pi_{F}(\omega,\vec{q})$ for $\vec{q}$ along the nesting vector direction $\vec{Q}_{SDW}$ in the Van Hove singularity region of energies $\varepsilon(\vec{k})$, i.e. for $q={Q}_{SDW}+\tilde{q}$ with $\tilde{q} \ll {Q}_{SDW}$ could be done along the energy axis with corresponding density of states near the Van Hove singularity $\nu$. Then, the integral $I_{\vec{q}}$ in the Eq. (\ref{Pci}) takes the form:
\begin{eqnarray}
I_{\vec{q}}=\nu\int_{-g_0+v_f\tilde{q}}^{g_0} {d\varepsilon}\sqrt{{g_0}^2-\varepsilon^{2}}\sqrt{{g_0}^2-(\varepsilon-v_f\tilde{q})^{2}}\approx \nu\left[\frac{1}{3}(v_f\tilde{q})^3-g_0(v_f\tilde{q})^2+\frac{4}{3}g^{3}_0\right]
\label{Pcd}
\end{eqnarray} 
\noindent Next, making analytic continuation to the axis of real frequencies: $i\omega\rightarrow \omega+i\delta, \delta \rightarrow +0$ in Eqs. (\ref{DC}), (\ref{Pci}) one finds the following expression for the retarded magnons Green's function $D^R(\omega,\vec{q})$, where a new notation $\omega+i\delta\equiv \tilde{\omega}$ is used for brevity:
\begin{eqnarray}
&D^R(\omega,\vec{q})=\dfrac{\omega^{2}_{0}(\vec{q})(4g_0^2-{\tilde{\omega}}^2)}{-{\tilde{\omega}}^{4}+{\tilde{\omega}}^{2}\left(4g_0^2+\omega^{2}_{0}(\vec{q})\right)+\omega^{2}_{0}(\vec{q})\left(E(\vec{q})-4g_0^2\right)}\,,\tilde{\omega}\equiv\omega+i\delta \,;\nonumber \\
&E(\vec{q})\equiv \dfrac{g^{2}_{bf}\tanh{\left(\dfrac{g_0}{2T}\right)}I_{\vec{q}}}{g_0}\,. \label{Dan}
\end{eqnarray}
\noindent The above result is remarkable: the poles of the $D^R(\omega,\vec{q})$ expression in Eq. (\ref{Dan}) provide two dispersion branches of magnons near the $\vec{Q}_{SDW}$ in the Brillouin zone:

\begin{eqnarray}
\omega_{\pm}(\vec{q})=\sqrt{\dfrac{\omega^{2}_{0}(\vec{q})+4g_0^2}{2}\pm \sqrt{\left(\dfrac{\omega^{2}_{0}(\vec{q})-4g_0^2}{2}\right)^2+\omega^{2}_{0}(\vec{q})E(\vec{q}) }}
\label{Bran}
\end{eqnarray}
\noindent When wave-vector of the magnons $\vec{q}$ is taken along the antiferromagnetic SDW nesting vector direction $\vec{Q}_{SDW}$ in the vicinity of the bottom of the antiferromagnetic excitations band, i.e. for $q={Q}_{SDW}+\tilde{q}$ with $\tilde{q} \ll {Q}_{SDW}$, one may approximate the bare magnon dispersion in the form: 
\begin{eqnarray}
\omega^{2}_{0}(\vec{q})=s^2\tilde{q}^2+\mu_0^2
\label{Bare}
\end{eqnarray}
\noindent introduced already in Eq. (\ref{Eu}), where $s$ is spin-wave velocity and $\mu_0$ is the 'magnon mass' characterising the short range character of antiferromagnetic fluctuations in the doped cuprates.  Substituting the above expression from Eq. (\ref{Bare}) into expression for the magnon dispersion branches in Eq. (\ref{Bran}) one finds dispersion plotted in Fig. \ref{MMM}, that follows famous experimental data for high-T$_c$ cuprates reported long ago \cite{Tranquada_2014}. 
\begin{figure}
\centering
    \includegraphics[width=0.8\linewidth]{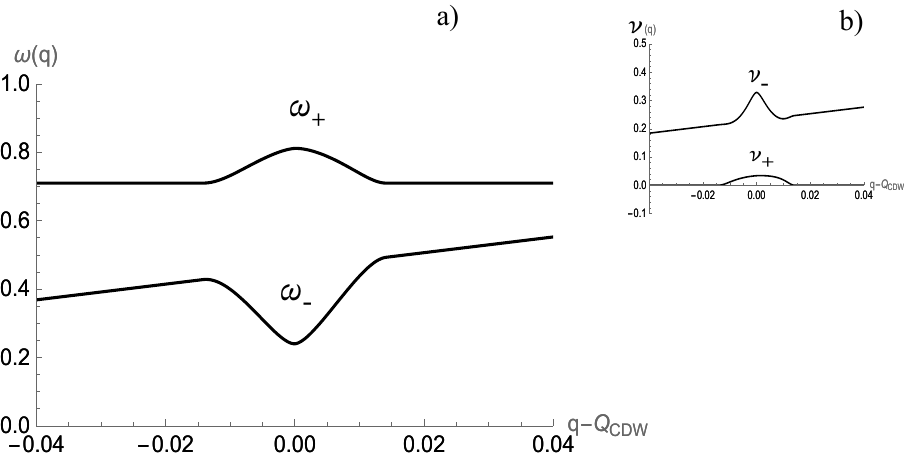}
\caption{a. Theoretical curves of the split phonons dispersion into hardened $\omega_{+}$ and softened $\omega_{-}$ brunches that follow from Eq. (\ref{Bran}); b) corresponding pole strengths $\nu_{\pm}=-2/\pi Im D^{R}(\omega_{\pm}, q)$ of the hardened $\omega_{+}$ and softened $\omega_{-}$ phonon modes in the vicinity of the nesting CDW wave-vector $Q_{CDW}$,  compare \cite{Egami}, Fig.4 left panel.}
\label{11}
\end{figure}
\noindent

\subsection{Splitting of the in-plane phonon dispersion brunches due to  CDW Q-balls scattering.}
Now, we turn to the splitting of the in-plane phonon dispersion brunches into the softened and hardened ones due to scattering of phonons on the CDW Q-balls close to the $Q_{CDW}$ wave vectors designated in Fig. \ref{QF}. Obtained theoretical curves of the phonon brunch splitting that follow from Eq. (\ref{Bran}) are presented in Fig. \ref{11}. Substituting the bare phonons dispersion:
 \begin{eqnarray}
\omega_{0}(\vec{q})=sq
\label{Bare_p}
\end{eqnarray} 
\noindent close to the 'nesting' wave-vectors $Q_{CDW}$ into the general expression in Eq. (\ref{Bran}) one obtains splitting of the in-plane phonon dispersion brunches  in the vicinity of these wave-vectors that was studied experimentally \cite{Egami}, compare Fig. \ref{11}a and Fig.4 in \cite{Egami}. The pole strength of the phonons Green's function in Eq. (\ref{Dan}) corresponding to the "hardened" dispersion brunch  $\omega_{+}(q)\approx 2g_0$ at $q\approx Q_{CDW}$ in Fig. \ref{11}a is small and corresponds to the excitation of the phonon by the virtual collapse of two fermions into the Cooper /local pairs condensate inside the Q-ball with transferring the energy twice the superconducting gap $2g_0$ to the lattice distortion, see $\nu_{+}$ line in the insert Fig. \ref{11}b. Simultaneously, the "softened" dispersion brunch:
 \begin{eqnarray}
\omega_{-}(q)\approx \omega_{0}(q)\sqrt{1-\dfrac{2E(q)}{4g_0^2-\omega^{2}_{0}(q)}}
\label{Soft}
\end{eqnarray} 
\noindent possesses maximum softening at $q\approx Q_{CDW}$, see Fig. \ref{11}a, due to the $\sim E(q)$ term defined in Eqs. (\ref{Pcd}), (\ref{Dan}).  The higher pole strength of this $\omega_{-}(q)$ brunch is plotted in the insert Fig. \ref{11}b, see $\nu_{-}$ line.  It is remarkable, that rather different dispersion curves in the figures Fig. \ref{MMM} and Fig. \ref{11} are obtained from one and the same Dyson equation (\ref{Bran}), but with the different bare dispersions  $\omega_{0}(\vec{q})$ in Eqs. (\ref{Bare}) and  (\ref{Bare_p}), and with the different Gor'kov's functions $F$, $F^{\dagger}$ that describe d-wave and s-wave Cooper-pairs condensates inside the SDW and CDW Q-balls respectfully.

\subsection{Dynamical structure factor $S(\vec{Q},\omega)$ of the Q-balls 'gas'}

\begin{figure}
\centering
    \includegraphics[width=0.8\linewidth]{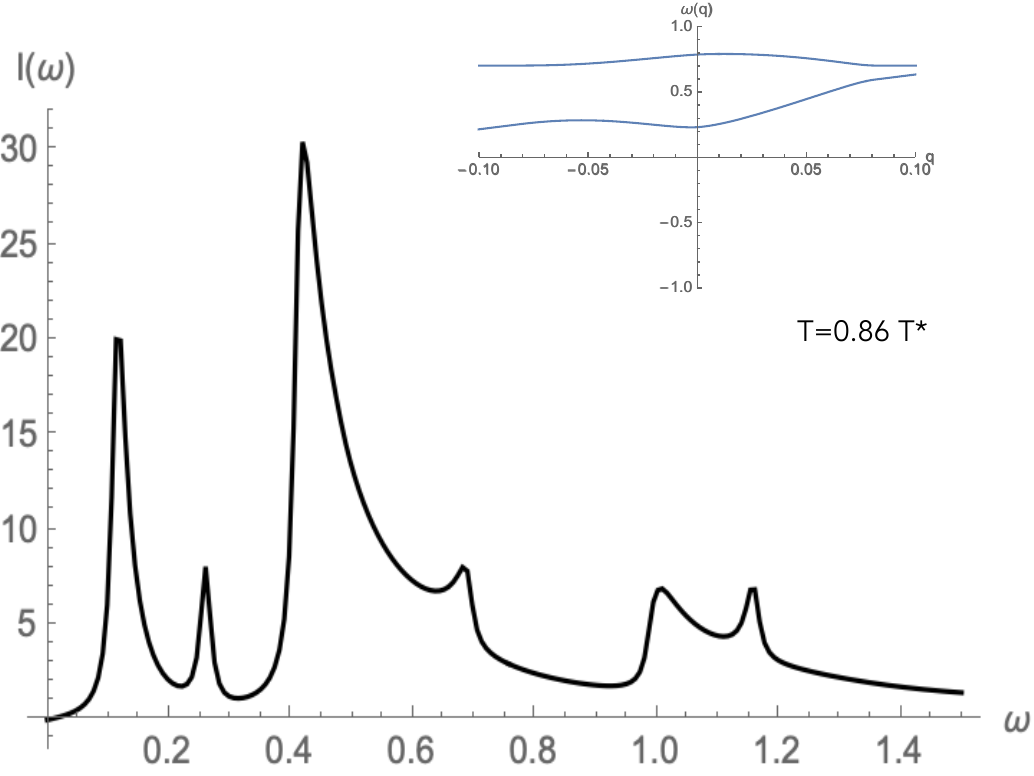}
\caption{Dynamic structure factor $S(Q_{CDW},\omega)\equiv I(\omega)$ is calculated from Eq. (\ref{MAPS0}) for temperature $T=0.86 T^*$, at which superconducting gap $g_0$ of Q-ball superconducting condensate reaches maximum in Eq. (\ref{g0star}), point $T_{c1}$ in Fig. \ref{1}. The insert represents theoretical curves of the split phonons dispersion into hardened $\omega_{+}$ and softened $\omega_{-}$ brunches that follow from Eq. (\ref{Bran}) at this same temperature $T=0.86 T^*$,  compare \cite{Egami}, Fig.3.}
\label{0}
\end{figure}
\noindent

\begin{figure}
\centering
    \includegraphics[width=0.8\linewidth]{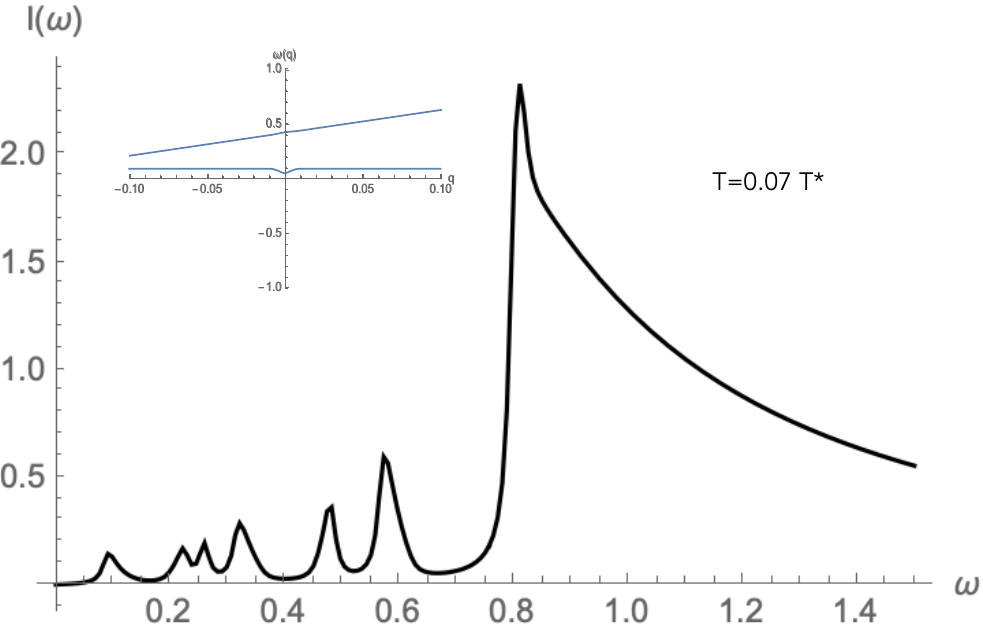}
\caption{Dynamic structure factor $S(Q_{CDW},\omega)\equiv I(\omega)$ is calculated from Eq. (\ref{MAPS0}) for temperature $T=0.07 T^*$, at which superconducting gap $g_0$ of Q-ball superconducting condensate approaches  minimum $g_0=0$ in Eq. (\ref{g0star}), point $T_{c2}$ in Fig. \ref{1}. The insert represents theoretical curves of the split phonons dispersion into hardened $\omega_{+}$ and softened $\omega_{-}$ brunches that follow from Eq. (\ref{Bran}) at this same temperature $T=0.07 T^*$.}
\label{00}
\end{figure}
\noindent
Since inelastic pulsed neutron scattering technique provides experimental data for the dynamical structure factor $S(\vec{Q},\omega)$, where $\vec{Q}$ is the momentum transfer and $\omega$ is the energy transfer \cite{Egami}, it is possible to compare experiment with the present theoretical results. Namely, $S(\vec{Q},\omega)$ is imaginary part of the density-density correlation function, that could be expressed via phonon Green's function found above in Eq. (\ref{Dan}) to the lowest order of the perturbation theory with respect to electron-phonon coupling in the following form:

\begin{eqnarray}
&S(\vec{Q},\Omega)=-\dfrac{2}{\pi}\Im{\Pi^{R}(\vec{Q},\Omega)}\,,\label{MAPS0}\\
&\Pi(\vec{Q},\Omega)=\langle n(\vec{r},\tau)n(\vec{r}\,^{ \prime} ,\tau')\rangle _{\vec{Q},\Omega}=T\int \dfrac{d\vec{q}}{(2\pi\hbar)^3}\sum_{\omega}D(\omega+\Omega,\vec{q}+\vec{Q})D(\omega,\vec{q})\,. \label{MAPS}
\end{eqnarray}
\noindent 
Substituting phonon Green's functions from Eq. (\ref{Dan}) and turning to retarded expression in Eq. (\ref{MAPS}) after analytical continuation to real frequency axis, one finds the dynamical structure factor $S(\vec{Q_{CDW}},\omega)\equiv I(\omega)$ at different temperatures indicated with two arrow bars $T_{c1,2}$ in the  Q-balls phase diagram in Fig. \ref{1}. Numerical results obtained from Eq. (\ref{MAPS0}) with superconducting gap $g_0$ from Eq. (\ref{g0star}) are presented in Figs. \ref{0} and \ref{00}.

\section{Conclusions}
\label{sec: fin}
To summarise, presented above theory of  the quantum orders in the form of periodic in the imaginary Matsubara time SDW/CDW fields signifies a new development beyond the Landau static mean-field order parameter approach.  Favourable comparison of the theory with experiment \cite{Niven, Li,Tranquada_2014, Egami, Keimer} indicates that  described above picture of free fermions outside the gas of Q-balls with Cooper pairs condensates inside of them below temperature T* opens an avenue for direct investigation of the thermodynamic quantum time crystals \cite{Efetov, Timur} of SDW/CDW dynamic quantum condensates forming space heterogeneous fluctuations \cite{Annette}. This realm of new phenomena possesses direct relation to observed physical properties of high-Tc superconductors. In a particular picture related with high-T$_c$ Q-balls scenario, the vanishing density of superconducting condensate at T* leads to inflation of Q-balls sizes, that self-consistently suppresses X-ray Bragg's peak intensity close to Q-ball phase transition temperature \cite{campi22}. Linear temperature dependence of electrical resistivity in the Q-balls phase due to scattering of electrons on the SDW/CDW condensates forming the Q-balls is also demonstrated. The T-linear dependence of electrical resistivity arises due to inverse temperature dependence of the Q-ball radius and linear dependences of SDW/CDW Q-ball amplitudes as functions of temperature in the "strange metal" phase. Simultaneously, the Cooper-pairs condensates inside the Q-balls give rise to diamagnetic response in the "strange metal" phase in accord with experiments \cite{Li, Zaanen}.
It is also demonstrated above that scattering of antiferromagnetic spin-waves (magnons) on the superconducting condensates of the Q-balls creates famous "hourglass" dispersion of magnetic excitations close to the antiferromagnetic wave-vectors in the Brillouin zone in accord with experiments \cite{Tranquada_2014}. Splitting of the in-plane phonon brunches into the softened and hardened ones close to Q$_{CDW}$ wave vectors in the Brillouin zone, as found in the inelastic pulsed neutron scattering experiments \cite{Egami}, is explained above by the scattering of phonons on the superconducting Q-ball condensates and is manifested by the calculated above phonons two-brunch Green's function $D(\omega,\vec{q})$ and dynamical structure factor $S(\vec{q},\omega)$, where $\vec{q}$ and $\omega$ is momentum and energy transfer respectively.   

\section{ACKNOWLEDGMENTS}
The author is grateful to prof. Antonio Bianconi for making available the experimental data on micro X-ray diffraction in high-T$_c$ cuprates prior to publication, to prof. Niven Barisic for presentation of major experimental data on the electronic  transport  properties in 'strange metal' phase, to prof. Jan Aarts for valuable discussion of  Q-ball CDW slide results, and to prof. Carlo Beenakker and his group for stimulating discussions during the whole work. This research was in part supported by  Grant No. K2-2022-025 in the framework of the Increase Competitiveness Program of NUST MISIS.


\begin{thebibliography}{99}
\bibitem{Mukhin(2022)} Mukhin, S.I.   Euclidean Q-Balls of Fluctuating SDW/CDW in the 'Nested' Hubbard Model of High-Tc Superconductors as the Origin of Pseudogap and Superconducting Behaviors. {\emph{Condens. Matter}} \textbf{2022}, \emph{7}, 31. https://doi.org/10.3390/condmat7020031

\bibitem{Mukhin(2022_1)}Mukhin, S.I.   Euclidean Q-balls of electronic spin/charge densities confining superconducting condensates as the origin of pseudogap and high-Tc superconducting behaviours. \emph{Ann. Phys.} \textbf{2022}, \emph{447}, 169000. https://doi.org/10.1016/j.aop.2022.169000. 

\bibitem{campuzano}K. Gofron, J.C. Campuzano, A.A. Abrikosov, M. Lindroos, A. Bansil, H. Ding, D. Koelling and B. Dabrowski. Observation of an "Extended" Van Hove Singularity in YBa$_2$Cu$_4$O$_{8}$  by Ultrahigh Energy Resolution Angle-Resolved Photoemission. {\emph{Phys. Rev. Lett.}}\textbf{1994}, \emph{73}, (24), 3302-3305.

\bibitem{Coleman (1985)} Coleman, S.R.   Q-balls. \emph{Nuclear Phys. B} \textbf{1985}, \emph{262}, 263--283.

\bibitem{Rosen}Rosen, G.   Particlelike Solutions to Nonlinear Complex Scalar Field Theories with Positive Definite Energy Densities. \emph{J. Math. Phys.} \textbf{1968}, \emph{9}, 996.  https://doi.org/10.1063/1.1664693.

\bibitem{Lee and Pang (1992)} Lee, T.D.; Pang, Y.   Nontopological solitons. \emph{Phys. Rept.} \textbf{1992}, \emph{221}, 251--350.

\bibitem{Niven}N. Barisic, Y. Li, G. Yu, X. Zhao, M. Dressel, A. Smontara, M. Greven.Universal sheet resistance and revised phase diagram of the cuprate high-temperature superconductors. {\emph{PNAS}} \textbf{2013}, \emph{110}, 12235.

\bibitem{Li} Li, L.; Wang, Y.; Komiya, S.; Ono, S.; Ando, Y.; Gu, G.D.; Ong, N.P.   Diamagnetism and Cooper pairing above Tc in cuprates. {\emph{Phys. Rev. B}} \textbf{2010}, \emph{81}, 054510.

\bibitem{Tranquada_2014}John M.Tranquada, Guangyong Xu, Igor A. Zaliznyak. Superconductivity, antiferromagnetism, and neutron scattering. {\emph{Journal of Magnetism and Magnetic Materials}}\textbf{2014}, \emph{350}, 148–160.

\bibitem{Egami} J.-H. Chung, T. Egami, R. J. McQueeney, {\it{et.al}}.  In-plane anisotropy and temperature dependence of oxygen phonon modes in YBa$_2$Cu$_3$O$_{6.95}$.{\emph{Phys. Rev. B}}\textbf{2003},\emph{67},  014517.

\bibitem{Keimer}M. Raichle, D. Reznik, D. Lamago, R. Heid, Y. Li, M. Bakr, C. Ulrich, V. Hinkov, K. Hradil, C.T. Lin, and B. Keimer. Highly anisotropic anomaly in the dispersion of the copper-oxygen bond-bending phonon in superconducting YBa$_2$Cu$_3$O$_7$ from inelastic neutron scattering. {\emph{Phys. Rev. Lett.}}\textbf{2011},\emph{107}, 177004.  https://doi.org/10.1103/PhysRevLett.107.177004.

\bibitem{Tranq1} Fradkin, E.; Kivelson, S.A.; Tranquada, J.M.   Colloquium: Theory of intertwined orders in high temperature superconductors. \emph{Rev. Mod. Phys.} \textbf{2015}, \emph{87}, 457.

\bibitem{MatMuk} Matveenko S.I. and Mukhin S.I. Pair density wave solution for a self-consistent model. {\emph{Phys. Rev. B}} \textbf{2025}, \emph{111}, 125155. https://doi.org/10.1103/PhysRevB.111.125155.

\bibitem{Pepin} Saheli Sarkar, Maxence Grandadam, and Catherine Pepin. Anomalous softening of phonon dispersion in cuprate superconductors.
{\emph{Physical Review Research}} \textbf{2021}, \emph{3}, 013162.

\bibitem{Seibold} Seibold, G.; Arpaia, R.; Peng, Y.Y.; Fumagalli, R.; Braicovich, L.; Di Castro, C.; Caprara, S.    Strange metal behaviour from charge density fluctuations in cuprates. \emph{Commun. Phys.} \textbf{2021}, \emph{4}, 1--6. https://doi.org/10.1038/s42005-020-00505-z

\bibitem{Caprara} Martina Fedele, Giacomo Merzoni, Marco Moretti Sala, Francesco Rosa, Nicholas B. Brookes, Floriana Lombardi, Sergio Caprara, Giacomo Ghiringhelli, and Riccardo Arpaia. Electron - phonon coupling revealed by charge density fluctuations in cuprate superconductors. arXiv:2602.18112v1. https://doi.org/10.48550/arXiv.2602.18112 .

\bibitem{Mukhin(2023)}	 Mukhin, S.I.  Possible Manifestation of Q-Ball Mechanism of High-Tc Superconductivity in X-ray Diffraction. Condens. Matter,\emph{8}, 16 (2023), https://doi.org/10.3390/condmat8010016.

\bibitem{campi22} Campi, G.;  Barba, L.; Zhigadlo, N.D.; Ivanov, A.A.; Menushenkov, A.P.; Bianconi, A.  
Q-Balls in the pseudogap phase of Superconducting HgBa2CuO4+y. \emph{Condens. Matter} \textbf{2023}, \emph{8}, 15. https://doi.org/10.3390/condmat8010015.

\bibitem{campi} Campi, G.; Bianconi, A.; Poccia, N.; Bianconi, G.; Barba, L.; Arrighetti, G.; Innocenti, D.; Karpinski, J.; Zhigadlo, N.D.; Kazakov, S.M.; et al. Inhomogeneity of charge-density-wave order and quenched disorder in a high-Tc superconductor. {\emph{Nature}} \textbf{2015}, \emph{525},  359--362.

\bibitem{Annette}Annette Bussmann-Holder, Jurgen Haase, Hugo Keller, Reinhard K. Kremer, Sergei I. Mukhin, Alexey P. Menushenkov, Andrei Ivanov, Alexey Kuznetsov, Victor Velasco, Steven D. Conradson, Gaetano Campi, and Antonio Bianconi. Nanoscale Lattice Heterostructure in High-Tc Superconductors. {\emph{Condens. Matter}} \textbf{2025}, \emph{10}, 56. https://doi.org/10.3390/condmat10040056 


\bibitem{Efetov} Konstantin B. Efetov. Mean-field thermodynamic quantum time-space crystal: Spontaneous breaking of time-translation symmetry in a macroscopic fermion system. Phys. Rev. B 100, 245128 (2019). https://link.aps.org/doi/10.1103/PhysRevB.100.245128 .
\bibitem{Timur}S. I. Mukhin and T. R. Galimzyanov. Classes of metastable thermodynamic quantum time crystals. {\emph{Phys. Rev. B}}\textbf{2019}, \emph{100}, 081103(R).






\bibitem{Egami1} R. J. McQueeney, Y. Petrov, T. Egami, M. Yethiraj, G. Shirane, and Y. Endoh. Anomalous Dispersion of LO Phonons in La$_{1.85}$Sr$_{0.15}$CuO$_4$ at Low Temperatures.{\emph{Phys. Rev. Lett.}}\textbf{1999},\emph{82}, (3), 628-631. https://doi.org/10.1103/PhysRevLett.82.628 .







\bibitem{Mukhin (2018)} Mukhin, S.I.  Negative Energy Antiferromagnetic Instantons Forming Cooper-Pairing Glue and Hidden Order in High-Tc Cuprates. \emph{Condens. Matter} \textbf{2018}, \emph{3}, 39.

\bibitem{BCS}Bardeen, J., Cooper, L.N. and Schriffer, J.R. Microscopic Theory of Superconductivity. {\emph{Physical Review}}, \textbf{1957}, \emph{106}, 162-164. (1957).

\bibitem{elis}  Eliashberg, G.M.  Interactions between electrons and lattice vibrations in a superconductor.  {\emph{JETP}} \textbf{1960}, \emph{11}, 696--702.

\bibitem{Chubukov (2003)} Abanov, A.; Chubukov, A.V.; Schmalian, J., Quantum-critical theory of the spin-fermion model and its application to cuprates: Normal state analysis.  \emph{Adv. Phys.} \textbf{2003}, \emph{52}, 119--218.

\bibitem{Bianconi (1994)} Bianconi, A.; Missori, M.    The instability of a 2D electron gas near the critical density for a Wigner polaron crystal giving the quantum state of cuprate superconductors. \emph{Solid State Commun.} \textbf{1994}, \emph{91}, 287--293. 

\bibitem{Louk}The author is grateful to Dr. Louk Rademaker for bringing attention to this matter.

\bibitem{Alex} A.A. Abrikosov, Fundamentals of the theory of metals. Elsevier Science Publishers B.V., P.O. Box 103 1000 AC Amsterdam, The Netherlands \textbf{1988}.

\bibitem{AGD} Abrikosov, A.A.; Gor'kov, L.P.; Dzyaloshinski, I.E. \emph{Methods of Quantum Field Theory in Statistical Physics}; Dover Publications: New York, NY, USA, 1963.

\bibitem{Witteker}Witteker, E.T.,Watson, G.N., A Course of Modern Analysis; Cambridge University Press: Cambridge, UK, 1996.

\bibitem{Derrick}Derrick, G.H.  Comments on nonlinear wave equations as models for elementary particles. \emph{J. Math. Phys.} \textbf{1964}, \emph{5}, 1252--1254.

\bibitem{Fischer}Renner, Ch. and Fischer, O. Vacuum tunneling spectroscopy and asymmetric density of states of ${\mathrm{Bi}}_{2}$${\mathrm{Sr}}_{2}$${\mathrm{CaCu}}_{2}$${\mathrm{O}}_{8+\mathrm{\ensuremath{\delta}}}$.  \emph{Phys. Rev. B}\textbf{1995}, \emph{51}, 9208-9218. https://link.aps.org/doi/10.1103/PhysRevB.51.9208.

\bibitem{Tranq0}J. M. Tranquada, H. Woo, T. G. Perring, H. Goka, G. D. Gu, G. Xu, M. Fujita and K. Yamada. Quantum magnetic excitations from stripes in copper oxide superconductors. \emph{Nature} \textbf{2004}, \emph{429}, 534--538.

\bibitem{Jan}Jan Zaanen, Planckian dissipation, minimal viscosity and the transport in cuprate strange metals, SciPost Phys. 6, 061 (2019). https://doi.org/10.21468/SciPostPhys.6.5.061.

\bibitem{Hinkov}V. Hinkov, D. Haug, B. Fauque, P. Bourges, Y. Sidis, A. Ivanov, C. Bernhard, C. T. Lin, and B. Keimer. Electronic liquid crystal state in the high-temperature superconductor YBa$_2$Cu$_3$O$_{6.45}$. Science 319, 597 (2008). DOI: 10.1126/science.1152309.

\bibitem{Zaanen} Keimer, B.; Kivelson, S.; Norman, M.; Uchida, S.; Zaanen, J.   From quantum matter to high-temperature superconductivity in copper oxides. \emph{Nature} \textbf{2015},  \emph{518}, 179.  https://doi.org/10.1038/ nature14165.

\bibitem{LL9}L. D. Landau, E. M. Lifshitz, L. P. Pitaevskii  Statistical Physics, Part 2, Vol. 9 (3rd ed.), Butterworth-Heinemann (1980), ISBN 0-7506-2636-4. 

\bibitem{Uemura} Uemura, Y.J.; Luke, G.M.; Sternlieb, B.J.; Brewer, J.H.; Carolan, J.F.; Hardy, W.;  Yu, X.H.  Universal correlations between Tc and n$_s$/m* in high-Tc cuprate superconductors. \emph{Phys. Rev. Lett.} \textbf{1989}, \emph{62}, 2317--2320.  

\bibitem{Egami22} R. J. McQueeney, Y. Petrov, T. Egami, M. Yethiraj, G. Shirane, and Y. Endoh, Anomalous Dispersion of LO Phonons in
La1.85Sr0.15CuO4 at Low Temperatures, Phys. Rev. Lett. 82, 628 (1999).
\bibitem{Reznik22} M. Raichle, D. Reznik, D. Lamago, R. Heid, Y. Li, M. Bakr, C. Ulrich, V. Hinkov, K. Hradil, C. T. Lin, and B. Keimer, Highly Anisotropic Anomaly in the Dispersion of the Copper-Oxygen Bond-Bending Phonon in Superconducting YBa2Cu3O7 from Inelastic Neutron Scattering, Phys. Rev. Lett. 107, 177004 (2011).

\end{thebibliography}
\end{document}